\documentclass[a4paper,11pt]{article}
\pdfoutput=1
\usepackage{jheppub}
\usepackage{cancel}
\usepackage{subcaption}
\usepackage{mathrsfs}
\usepackage{amsmath}   % for \overset
\usepackage{feynmp-auto,expdlist}
\usepackage{float}

\newcommand{\beq}[1]{\begin{equation}\label{#1}}
 \newcommand{\eeq}{\end{equation}}
 \newcommand{\bea}[1]{\begin{eqnarray}\label{#1}}
 \newcommand{\eea}{\end{eqnarray}}

\title{Quantum Entanglement of Circular Strings as a Probe for Topologically Charged Spacetimes}
 \author{Ai-chen Li $^{a,b}$, Xin-Fei Li $^{a}$, Xuanting Ji $^{c}$}
 \affiliation[a]{School of Science, Guangxi University of Science and Technology, 545026 Liuzhou, China} 
\affiliation[b]{Departament de F\'{i}sica Qu\`{a}ntica i Astrof\'{i}sica, Institut de Ci\`{e}ncies del Cosmos (ICCUB), Universitat de Barcelona, Mart\'{i} i Franqu\`{e}s 1, E-08028 Barcelona, Spain} 
\affiliation[c]{Department of Applied Physics, College of Science, China Agricultural University, Beijing 100083, China}

\emailAdd{alexkenlee@163.com}
\emailAdd{xfli@gxust.edu.cn}
\emailAdd{jixuanting@cau.edu.cn}

\abstract{Motivated by the limited understanding of entanglement entropy in non-asymptotically AdS spacetimes, we develop a framework in which a circular string is embedded as a quantum probe in a spherically symmetric curved spacetime, and its quadratic fluctuations are quantized using the squeezed-state formalism. This construction naturally yields two-mode quantum states and the associated von Neumann entropy, providing a direct measure of particle–antiparticle entanglement. The resulting entanglement serves as an effective probe of the underlying geometry, granting access to intrinsic features that are not readily captured by classical observables such as geodesic motion. As a concrete application, and as representative toy models of spacetimes with topological defects, including wormhole geometries, we investigate backgrounds with topological charge, focusing on global monopole and monopole wormhole configurations. We show that the entanglement generated by the probe string exhibits a clear qualitative distinction between these backgrounds and is highly sensitive to the global structure of the spacetime, in particular to the deficit angle. These results illustrate the utility of quantum correlations as diagnostic tools for probing geometric properties beyond the classical regime and offer a complementary perspective on the interplay between spacetime structure and quantum entanglement.}

\begin{document} 
\maketitle
\flushbottom

\section{Introduction}
Over the past two decades, quantum entanglement has evolved from a cornerstone of quantum information science \cite{Nielsen_Chuang_2010} into a fundamental tool for exploring quantum gravity \cite{Ryu:2006bv, Hubeny:2007xt, Faulkner:2013ana, Liu:2025krl}, cosmology \cite{Brahma2020, Belfiglio:2025cst,Tejerina-Perez:2024opu,Tejerina-Perez:2026iaw,Tejerina-Perez:2026yrq}, and strongly correlated systems in condensed matter physics \cite{Amico2008}. Within this framework, entanglement entropy—which quantifies the degree of entanglement between subsystems—has proven to be a particularly insightful quantity. It not only unveils the intricate structure of quantum states but also establishes a profound connection between spacetime geometry and quantum field theory \cite{Hartnoll:2015fca, Frenkel:2021yql}. A paradigmatic example of this connection is the holographic principle, where boundary entanglement entropy is dual to bulk geometric quantities \cite{ Ryu:2006ef, Lewkowycz:2013nqa, Engelhardt:2014gca}.

In addition, the squeezed-state formalism \cite{Grishchuk:1989ss,Grishchuk:1990bj,Albrecht:1992kf} provides a particularly natural and powerful framework for characterizing the quantum, or more precisely semiclassical, features of systems subject to gravitationally induced time dependence, especially when the dynamics is governed by a quadratic-order perturbative action \cite{Das:2025kyn}. Such situations commonly arise when the background spacetime exhibits explicit time dependence, as in cosmological settings \cite{Kanno:2021vwu,Martin:2007bw,Haque:2020pmp,Bhattacharyya:2020rpy,Zhai:2024tkz,Li:2024kfm,Liu:2021nzx,Adhikari:2022oxr} or during highly dynamical processes such as black-hole mergers \cite{Kanno:2025how}. Even in flat or stationary spacetimes, time-dependent matter sources \cite{Ando:2020kdz}, when treated without back-reaction, generically lead to squeezed quantum states once the action is expanded to second order in either metric or matter perturbations. 

The development of the AdS/CFT correspondence has further elevated entanglement entropy  to a central role, particularly through the Ryu–Takayanagi  prescription \cite{Ryu:2006bv, Ryu:2006ef} and its covariant generalizations \cite{Hubeny:2007xt, Lewkowycz:2013nqa, Dong:2016hjy}, where it serves as a standard diagnostic for probing the geometric and dynamical properties of asymptotically AdS spacetimes \cite{Casini:2011kv, Li:2025ksw, Auzzi:2025sep}, and also as a fundamental block to construct spacetime \cite{Swingle:2009bg, VanRaamsdonk:2010pw, Freedman:2016zud}. By contrast, for non-asymptotically AdS geometries, the role of entanglement entropy remains much less explored. In this work, we argue that a complementary and broadly applicable framework can be constructed by introducing lower-dimensional world-sheet probes into such backgrounds. In particular, by embedding extended objects, such as strings, into non-AdS spacetimes and quantizing their quadratic fluctuations using the squeezed-state formalism, one naturally obtains two-mode quantum states associated with the perturbative degrees of freedom. This construction provides direct access to the corresponding von Neumann entropy, which characterizes the particle–antiparticle entanglement generated by the quantum fluctuations. In this way, entanglement entropy derived from quantum probes can serve as an alternative diagnostic tool for extracting physical information about non-AdS spacetimes, thereby extending the scope of entanglement-based approaches beyond the holographic setting. 

Against this backdrop, the present work investigates a circular cosmic string propagating in a class of spherically symmetric, stationary spacetimes endowed with nontrivial topological charge, namely the global monopole and monopole–wormhole geometries \cite{Nascimento:2019qor}, which exhibit a rich variety of intriguing physical phenomena \cite{Aounallah:2020rlf,Furtado:2020puz,Soares:2021uep,Kanesom:2021ytb,Pereira:2023xtb,Pereira:2023vud,Mustafa:2024cdu,Toshmatov:2024wpn,Mustafa:2024htt,Ahmed:2024eft,Li:2026cix}. We focus on the transverse quantum fluctuations of the probe string at the level of the quadratic-order perturbative action. Employing the squeezed-state formalism for canonical quantization, we construct the corresponding two-mode quantum states generated by the quadratic Hamiltonian, in close analogy with particle production in time-dependent backgrounds such as an expanding universe \cite{Parker1969, Birrell_Davies_1982}. Based on these quantum states, we further compute the associated particle–antiparticle entanglement entropy. Our aim is to use the entanglement induced by quantum fluctuations of the probe string as a diagnostic to distinguish, from the perspective of quantum probes, the underlying physical differences between pure global monopole spacetimes and monopole wormhole geometries. This provides a complementary viewpoint beyond traditional classical analyses, such as geodesic motion \cite{Shi:2016bxz,Cruz:2004ts} and gravitational lensing \cite{Soares:2023err}. The present study thus offers an analytically tractable framework for probing quantum information in curved spacetimes and connects to broader ideas relating spacetime geometry and entanglement, in particular the spirit of the ER=EPR conjecture \cite{Maldacena:2013xja,Susskind:2014yaa}. It may also have potential implications for the phenomenology of cosmic strings \cite{Izquierdo:2024jfw}.

Building on the above analysis and motivations, the structure of this work is organized as follows. In Sec.\ref{ReviewGMWH}, we present a detailed review of a class of generic spherically symmetric and stationary spacetimes carrying a topological charge, arising within the framework of Eddington-inspired Born–Infeld (EiBI) gravity. In particular, we focus on the subset of solutions encompassing both global monopole and monopole wormhole (GM/WH) geometries. Subsequently, in Sec.~\ref{ProbeStringQuadraPer}, we derive the dynamics of a circular cosmic string propagating in these GM/WH backgrounds and construct the corresponding quadratic-order perturbative action describing small fluctuations around the classical string trajectory. In Sec.~\ref{SqueeCanoQuantize}, we then present a detailed canonical quantization of the quadratic string perturbations using the squeezing formalism. In this context, we derive explicit expressions for the squeezing parameters in terms of the mode functions and systematically construct the associated time-evolution operators based on the quadratic Hermitian Hamiltonian. By applying the time-evolution operator to the Fock space constructed from the initial vacuum state, we obtain the corresponding two-mode quantum states that describe the quantum excitations of the string fluctuations in Sec.~\ref{QuantumStateAndEntropy}. Based on these states, we further construct the associated von Neumann entropy, which provides a natural measure of the particle–antiparticle entanglement generated by the quantum fluctuations of the probe string during its periodic motion. Moreover, supported by numerical analysis, we demonstrate that the entanglement associated with the string fluctuations exhibits a pronounced qualitative distinction between monopole spacetimes with and without a wormhole structure. In particular, the dependence of the entanglement entropy on the deficit angle factor $\alpha_0 = 1 - 8\pi G \eta^2$ differs significantly between the two cases, highlighting the sensitivity of quantum correlations to the underlying global geometry. Finally, we conclude in Sec.~\ref{ConAndDiscu} with a discussion of the physical implications of our results and outline potential future directions, especially regarding their relevance to the holographic characterization of non-asymptotically AdS spacetimes.

\section{A topologically charged monopole/wormhole  from Eddington-inspired Born-Infeld (EiBI) gravity }\label{ReviewGMWH}

\subsection{Eddington-inspired Born-Infeld (EiBI) gravity}
In this section we briefly review on EiBI gravity in the spirit of Ref. \cite{Nascimento:2019qor}. Further, by considering the canonical K-essence matter, the spacetime element of the wormhole/global monopole are solved.

EiBI gravity is a metric-affine theory, whose action is 
\begin{eqnarray}
S_{\mathrm{EiBI}} = \frac{1}{8\pi G \epsilon} \int d^4x \left( \sqrt{-|g_{\mu\nu} + \epsilon R_{(\mu\nu)}(\Gamma)|} - \lambda \sqrt{-|g_{\mu\nu}|} \right)+ S_m[g_{\mu\nu}, \Phi],
\end{eqnarray}
where $G$ is Newton's gravitational constant, $\epsilon$ is a parameter with dimension of area that relates to the nonlinearity of the theory, $\Gamma$ is an independent connection, and $\lambda = 1 + \epsilon \Lambda$ with $\Lambda$ the cosmological constant (vertical bars denote matrix determinant). For simplicity, we choose $\lambda = 1$. The variation of the action above with respect to the metric $g_{\mu\nu}$ and the connection $\Gamma^\alpha_{\mu\nu}$ leads to field equations 
\begin{equation}\label{EoM1}
	\sqrt{-h}h^{\mu\nu}=\sqrt{-g}\left(g^{\mu\nu}-\epsilon\kappa^2T^{\mu\nu}\right),
\end{equation}
\begin{equation}\label{EoM2}
	\nabla_{\mu}\left(\sqrt{-h}h^{\alpha\beta}\right)=0.
\end{equation}
The auxiliary metric $h_{\mu\nu}$ relates to the physical metric $g_{\mu\nu}$ by means of a deformation matrix $\bf{\Omega}^{\alpha}_{\phantom{\alpha}\nu}$ according to
\begin{equation}\label{h-conformal-g}
	h_{\mu\nu}=g_{\mu\alpha}\boldsymbol{\Omega}^{\alpha}_{\phantom{\alpha}\nu}, \quad h^{\mu\nu}=\left(\boldsymbol{\Omega}^{-1}\right)^{\mu}_{\phantom{\mu}\alpha}g^{\alpha\nu}.
\end{equation}
Substituting the relations above in Eq. \eqref{EoM1}, one obtains
\begin{equation}\label{Omega}
	\sqrt{|\boldsymbol{\Omega}|}\left(\boldsymbol{\Omega}^{-1}\right)^{\mu}_{\phantom{\mu}\nu}=\delta^{\mu}_{\phantom{\mu}\nu}-\epsilon\kappa^2T^{\mu}_{\phantom{\mu}\nu} \ , 
\end{equation}
which indicates that  the deformation is determined by the local stress-energy densities. 

In order to achieve the deformation in Eq. \eqref{Omega}, one needs to assume the matter source of  an anisotropic fluid with the form that
 \begin{equation}\label{tensor}
T^{\mu}_{\phantom{\mu}\nu}=\text{diag} \left(-\rho,-\rho, P_{\theta},P_{\theta}\right).
 \end{equation}
 Given the diagonal form of 
 \begin{eqnarray}
 \boldsymbol{\Omega}^{\mu}_{\nu} = \mathrm{diag}\left(\boldsymbol{\Omega}_+, \boldsymbol{\Omega}_+, \boldsymbol{\Omega}_-, \boldsymbol{\Omega}_-\right),
 \end{eqnarray}
 where
\begin{equation}\label{omegapm}
\boldsymbol{\Omega}_{-}=1+\epsilon\kappa^2 \rho,\qquad\boldsymbol{\Omega}_{+}=1-\epsilon\kappa^2 P_{\theta}.
\end{equation}
are derived directly  from Eq. \eqref{Omega}.
For a static, spherically symmetric spacetime, we start with a general ansatz for the auxiliary metric $h_{\mu\nu}$,
\begin{align}
d\tilde{s}^{2}= -A(x)e^{2\Phi (x)}dt^{2} + \frac{dx^{2}}{A(x)}&+\tilde{r}^{2}(x)(d\theta^{2} + \sin^{2}\theta d\phi^{2}). \label{ds2tild1}
\end{align}
Calculating the Ricci tensor $R^{\mu}_{\nu}(h)$ for this ansatz yields
\begin{align}
& R^{t}_t-R^{x}_x = \frac{2}{\tilde{r}}\left(\frac{d^{2}\tilde{r}}{dx^{2}} -\frac{d\Phi}{dx}\frac{d\tilde{r}}{dx}\right), \label{Rtx} \\
& R^{\theta}_{\theta} = \frac{1}{\tilde{r}^{2}}\left[1 - \tilde{r}\frac{d\tilde{r}}{dx}\left(A\frac{d\Phi}{dx} +\frac{dA}{dx}\right) - A\left(\tilde{r}\frac{d^{2}\tilde{r}}{dx^{2}} +\left(\frac{d\tilde{r}}{dx}\right)^{2}\right)\right]. \label{Rtheta}
\end{align}
From the matter source we will consider, it follows that $T^{t}_t=T^{x}_x$. Eq. \eqref{Rtx} then implies $\left(\frac{d^{2}\tilde{r}}{dx^{2}} -\frac{d\Phi}{dx}\frac{d\tilde{r}}{dx}\right) = 0$. Without loss of generality, we can use the residual gauge freedom to set $\Phi(x) = 0$ and $\tilde{r} = x$. Consequently, the line element for $h_{\mu\nu}$ simplifies to
\begin{equation}
d\tilde{s}^{2} = -A(x)dt^{2} + \frac{dx^{2}}{A(x)} + x^{2}(d\theta^{2} + \sin^{2}\theta d\phi^{2}). \label{ds2tild12}
\end{equation}
The physical metric $g_{\mu\nu}$ is related to $h_{\mu\nu}$ via the deformation matrix $\boldsymbol{\Omega}^{\mu}_{\nu}$, i.e., $h_{\mu\nu}=g_{\mu\alpha}\boldsymbol{\Omega}^{\alpha}_{\nu}$. Then, the line element for $g_{\mu\nu}$ can be directly written as
\begin{equation}
ds^{2} = -\frac{A(x)}{\boldsymbol{\Omega}_+} dt^{2} + \frac{1}{\boldsymbol{\Omega}_+ A(x)} dx^{2} + r^{2}(x)(d\theta^{2} + \sin^{2}\theta d\phi^{2}), \label{ds2}
\end{equation}
where the physical radial coordinate is given by $r^{2}(x) = \frac{x^{2}}{\boldsymbol{\Omega}_-}$. This relation, combined with the expressions for $\boldsymbol{\Omega}_\pm$ in Eq. \eqref{ds2}, leads to $x^{2} = r^{2} + \epsilon \kappa^{2}r^{2}\rho$.

The component $R^{\theta}_{\phantom{\theta}\theta}(h)$  now becomes
 \begin{equation}\label{Rtheta}
 R^{\theta}_{\phantom{\theta}\theta}(h)=\frac{1}{x^2}\left(1-A-x\frac{dA}{dx}\right).
 \end{equation}
For  simplicity, let's take the ansatz that $A(x)=1-\frac{2M(x)}{x}$ where where $M(x)$ plays the role of a mass function, analogous to the Schwarzschild case. Then, Eq. \eqref{Rtheta} becomes
\begin{eqnarray}
     R^{\theta}_{\phantom{\theta}\theta}(h)=\frac{2}{x^2}\frac{dM}{dx}.
\end{eqnarray}
Further, substituting the $\theta$-component of \eqref{Omega} into the equation above, one yields to
\begin{eqnarray}
    \frac{dM(x)}{dx}=\frac{\kappa^2r^2\rho}{2}.
\end{eqnarray}
Hence, the field equations for the auxiliary metric yield
	\begin{equation} \label{Ax}
		A(x) = 1 - \frac{2M_0}{x} - \frac{\kappa^2}{x} \int r^2 \rho \, dx,
	\end{equation}
	where $M_0$ is an integration constant. In order to obtain the algebraic structure of the stress tensor in Eq.~\eqref{tensor}, i.e., $T_{\mu\nu} = \operatorname{diag}(-\rho, \rho, P_\theta, P_\theta)$, in the next subsection we introduce a class of nonlinear $\sigma$-models whose corresponding stress-energy tensor fits this structure.

\subsection{Spacetime of wormhole carrying a global monopole}
The general K-monopole action \cite{Babichev:2006cy} is
\begin{equation}\label{kmonopole}
	S=\int\left[\mathcal{K}(X)-\frac{\lambda}{4}(\vec{\Phi}\cdot\vec{\Phi}-\eta^2)^2\right]\sqrt{-g}d^4x,
\end{equation}
  where  $\vec{\Phi}\equiv\{\Phi^i\} (i=1,2,3)$ a triplet of coupled real scalar fields and  $\mathcal{K}(X)$ is generally a non-canonical kinetic term with $X=\frac{1}{2}\partial_\mu\vec{\Phi}\cdot\partial^\mu\vec{\Phi}$. The action \eqref{kmonopole} exhibits spontaneous symmetry breaking $O(3)\rightarrow O(1)$ with the symmetry breaking scale $\eta$. The constant $\lambda$ acts as a Lagrange multiplier enforcing  the constraint  $\vec{\Phi}^2=\eta^2$.
  
  On the vacuum manifold $S^2$ of the field $\vec{\Phi}$, one can introduce local coordinates $\phi^a$ ($a=1,2$) such that $\vec{\Phi}=\vec{\Phi}(\phi^a)$. 
The kinetic term then becomes
\begin{eqnarray}
    X = \frac{1}{2}\eta^2\,\zeta_{ab}\,\partial_\mu\phi^a\partial^\mu\phi^b,
\end{eqnarray}
where $\zeta_{ab} = \frac{\partial\vec{\Phi}}{\partial\phi^a}\cdot\frac{\partial\vec{\Phi}}{\partial\phi^b}$
is the metric on the target $S^2$.
 The field equation and the energy-momentum tensor are derived as in Ref. \cite{Prasetyo:2017rij}
 \begin{equation}\label{EoM}
 	\frac{1}{\sqrt{-g}}\partial_\mu\left[\sqrt{-g}\eta^2\mathcal{K}_X\zeta_{bi}\partial^{\mu}\phi^i\right]-\frac{\mathcal{K}_X}{2}\eta^2\partial_{\mu}\phi^i\partial^{\mu}\phi^j\frac{\partial\zeta_{ij}}{\partial\phi^b}=0,
 \end{equation}
 \begin{equation} \label{Tmonopole1}
T^{\mu}_{\phantom{\mu}\nu}=\delta^{\mu}_{\nu}\mathcal{K}-\eta^2\mathcal{K}_{X}\partial^{\mu}\phi^a\partial_{\nu}\phi^a.
 \end{equation}
where  $\mathcal{K}_X$ denotes a derivative with respect to $X$.

 For a spherically symmetric metric, that is, $ds^2=-A(r)dt^2+B(r)dr^2+C(r)(d\theta^2+\sin^2\theta d\varphi^2)$, the ansatz $\phi^1=\theta$ and $\phi^2=\varphi$ identically satisfies the field equations \eqref{EoM} when choosing that
 \begin{eqnarray}
 \text{diag}(\zeta_{ij})=(1, \text{sin}^2\theta),     
 \end{eqnarray}
 as in Ref. \cite{Gell-Mann:1984ddz}. With this chose, a direct calculation gives $X = \eta^2 / r^2$.  Further,  combing  eq. \eqref{Tmonopole1} with the ansatz above, the energy-momentum tensor yields the diagonal form 
\begin{eqnarray}\label{Tmonopole2}
    T^\mu_\nu = \mathrm{diag}\left(\mathcal{K},\mathcal{K},\mathcal{K}-X\mathcal{K}_X,\mathcal{K}-X\mathcal{K}_X\right).
\end{eqnarray}
	
For simplicity, we consider the canonical case $\mathcal{K}(X)=X$. Comparing eq. \eqref{Tmonopole1} and eq. \eqref{Tmonopole2}, one obtains 
\begin{eqnarray}
    \rho=\frac{\eta^2}{r^2}, \ \  P_{\theta}=0.
\end{eqnarray}
This corresponds to the energy density profile of a global monopole with charge $\eta$.

By the relations $r^2 = x^2 / \boldsymbol{\Omega}_-$ and $\boldsymbol{\Omega}_- = 1 + \epsilon \kappa^2 \rho$, we get
	\begin{equation}
		r^2 = x^2-\epsilon \kappa^2 \eta^2.
	\end{equation}
Inserting $\rho = \eta^2/r^2$ into eq. \eqref{Ax} 
	\begin{equation}
		A(x) = 1 - \frac{2M_0}{x}-\kappa^2 \eta^2. 
	\end{equation}
Finally, after substituting the value of $x$,
 the general solution of metric in terms of $r$ becomes
 \begin{eqnarray}\label{Ac}
 	ds^2&=&-f(r)dt^2 +\frac{r^2}{r^2+\epsilon \kappa^2\eta^2}f(r)^{-1}dr^2 \notag \\ &+&
	r^2\left(d\theta^2+\sin^2\theta d\phi^2\right),
 \end{eqnarray}
where $f(r)=1-\kappa^2\eta^2-\frac{2M_0}{\sqrt{r^2+\epsilon\kappa^2\eta^2}}$.  Note that gravitational lensing is able to interpolates  the global monopole  where $\epsilon>0$ and the wormhole $\epsilon<0$ as in Ref. \cite{Soares:2023err}. We emphasize that the generic metric above contains several known geometries as special cases. The branch with $\epsilon>0$ corresponds to the global monopole sector, while the branch with $\epsilon<0$ describes a monopole wormhole with a nonzero throat radius and a solid-angle deficit. In the zero-deficit limit $\alpha_0\to 1$, or equivalently $\eta\to 0$ with the throat radius kept fixed, the monopole wormhole branch reduces to the usual Ellis--Bronnikov wormhole. This connects the present GM/WH background with the standard global monopole spacetime~\cite{Barriola:1989hx}, the EiBI global-monopole and wormhole solutions~\cite{Banados:2010ix,Nascimento:2019qor,Lambaga:2018yzv}, and the topologically charged Ellis--Bronnikov-type wormhole studied through scalar fields, Klein--Gordon oscillators, and gravitational lensing~\cite{Aounallah:2020rlf,Furtado:2020puz,Soares:2023err}. In the following sections, we extend this line of investigation by using the particle--antiparticle entanglement of a quantized circular-string probe as a diagnostic of the background geometry.

By choosing $M_0=0$, this describes an Ellis-like wormhole with topological charge \cite{Aounallah:2020rlf}. Then, after absorbing the factor $1-\kappa^2\eta^2$ into time coordinate in Eq. \eqref{Ac},  and introducing $\epsilon=\pm L_0^2$ for further convenience, the spacetime element Eq. \eqref{Ac} becomes
\begin{align}
\label{GMorWHinEiBI}
&ds^2=-dt^2+\frac{dr^{2}}{\left(1-\kappa^{2}\eta^{2}\right)\left(1\pm\frac{L_{0}^{2}\kappa^{2}\eta^{2}}{r^{2}}\right)}+r^2\left(d\theta^2+\sin^2\theta d\varphi^2\right). 
\end{align}
where $+$ sign denotes the case for a global monopole, and $-$ sign denotes the case for  a topologically charged wormhole which has a solid angle deficit compared with the Ellis-Bronnikov wormhole. For the global monopole,  it is a simply connected spacetime, and geodesics terminate at $r=0$. 
For $\epsilon<0$, the coordinate $r$ belongs to $[r_{\text{min}},\infty )$, where $r_{\text{max}}=L_0 \kappa \eta$, which represents the radius of throat the wormhole, connecting two asymptotic regions.
This spacetime returns Morris-Thorn wormhole as $\kappa \eta \to 0$. In summation, their differences lie in global topology, singularity structure, causal structure and geodesic completeness, which are entirely determined by the sign of the EiBI parameter $\epsilon$. 

Despite distinct geometries of the global monopole and the wormhole, they share three key commonalities: 
(1) Both are supported by matter that obeys the energy conditions \cite{Nascimento:2019qor, Lambaga:2018yzv}; 
(2) In the asymptotic limit $r \to \infty$, the deficit angle is $2\pi(1-\kappa^2\eta^2)$, which has been analyzed in both weak and strong field regimes via gravitational lensing effects \cite{Soares:2023err}; 
(3) Both solutions are derived from the EiBI gravity action coupled to the k-monopole matter sector, with the sign of the parameter $\epsilon$ dictating the emergence of the corresponding geometric phase.

For simplification in the following context, we also take some abbreviations $\alpha_{0}^{2}=1-\kappa^{2}\eta^{2}$ and $c_1=\pm1\,,\, r_{0}=L_{0}\kappa\eta$. 

\section{Quadratic perturbations of the circular string \label{ProbeStringQuadraPer}}

\begin{figure}[ht]
 \begin{center}
    \includegraphics[width=\columnwidth]{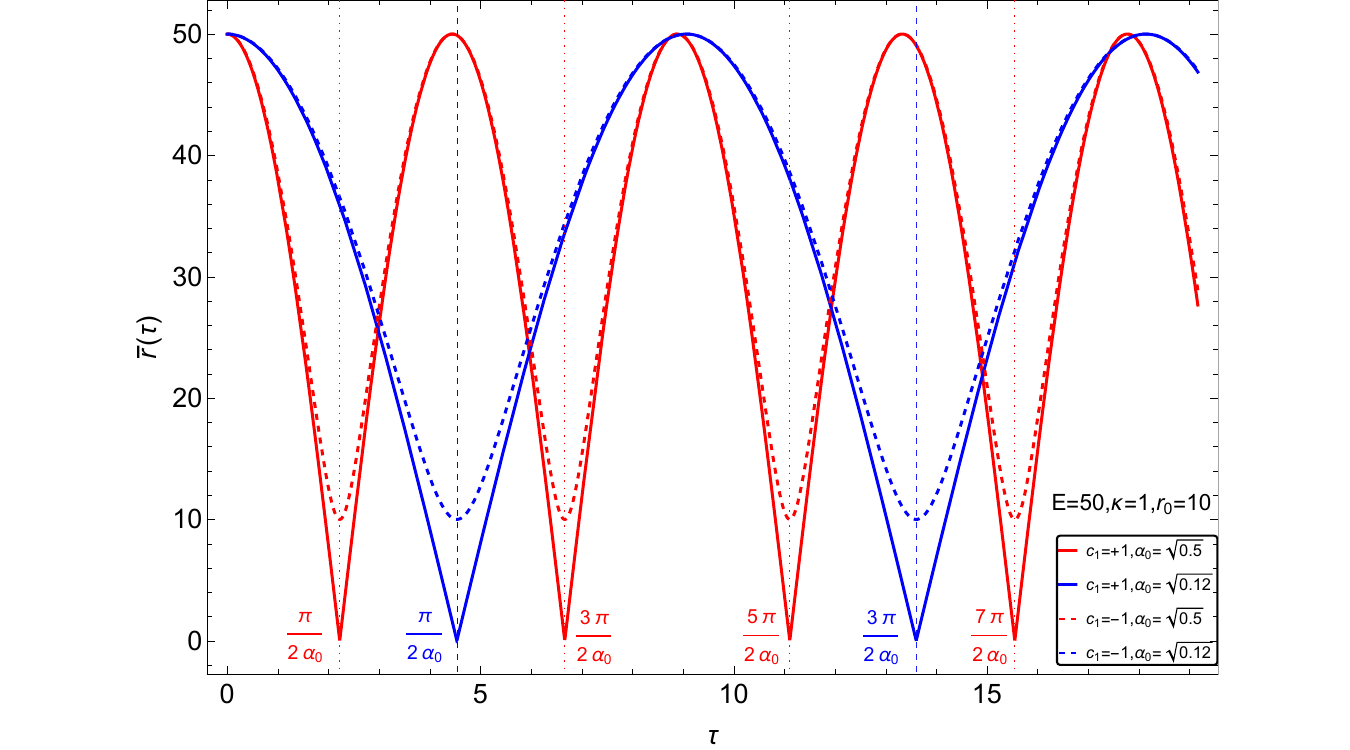}
 \caption{Classical trajectories of the probe circular cosmic string in the global monopole spacetime ($c_1=+1$, solid line) and in the monopole wormhole spacetime ($c_1=-1$, dashed line) as functions of $\tau$, for representative parameter choices. Note that, in the wormhole case, the probe string reaches a minimal radius corresponding to the throat position $r_0\neq 0$, whereas in the monopole case it can reach the origin $r_0=0$.}
 \label{BackStringGMWH}
 \end{center}
 \end{figure}

In this section, we closely follow the technical framework developed in Ref.~\cite{Li:2026oxy}; accordingly, many intermediate derivations are not presented here. We begin with the Polyakov action describing a relativistic string,
\begin{align}
\label{BackStringAction}
&S_{(0)}=\frac{-1}{4\pi\alpha^{\prime}}\int d\tau d\sigma\sqrt{-h}h^{\mathtt{A}\mathtt{B}}G_{\mathtt{A}\mathtt{B}},
\end{align}where $h_{\mathtt{A}\mathtt{B}}$ and $\alpha^\prime$ denote the intrinsic world-sheet metric and the string tension, respectively, while $G_{\mathtt{A}\mathtt{B}}$ denotes the induced metric,
\begin{align}
\label{InduceMetric}
&G_{\mathtt{A}\mathtt{B}}=g_{\mu\nu}\frac{\partial x^{\mu}}{\partial\xi^{\mathtt{A}}}\frac{\partial x^{\nu}}{\partial\xi^{\mathtt{B}}}\,,\,\xi^{\mathtt{A}}=\{\tau,\sigma\}\,,\,x^{\mu}=\{t,r,\theta,\phi\},
\end{align}
which are obtained from the spacetime embedding. By varying \eqref{BackStringAction} with respect to the world-sheet metric $h_{\mathtt{A}\mathtt{B}}$ and the embedding coordinates $x^\mu(\xi)$, we obtain the corresponding equations of motion together with the associated constraint equations,
\begin{align}
\label{EOMsBackEmbedding}
0&=h^{\mathtt{A}\mathtt{B}}\left(\nabla_{\mathtt{A}}\nabla_{\mathtt{B}}x^{\mu}+\Gamma_{\alpha\beta},^{\mu}\frac{\partial x^{\alpha}}{\partial\xi^{\mathtt{A}}}\frac{\partial x^{\beta}}{\partial\xi^{\mathtt{B}}}\right),\\
\label{RestrictBackEmbedding}
 0&=\frac{1}{2}h_{\mathtt{A}\mathtt{B}}h^{\mathtt{A}_{1}\mathtt{B}_{1}}G_{\mathtt{A}_{1}\mathtt{B}_{1}}-G_{\mathtt{A}\mathtt{B}}.
\end{align}Without loss of generality, the worldsheet metric can be gauge-fixed to $h_{\mathtt{A}\mathtt{B}}=\eta_{\mathtt{A}\mathtt{B}}$ by exploiting the conformal symmetry of the Polyakov action. To describe a circular string with a time-dependent radius, confined to the equatorial plane of a static spacetime with spherical topology, we adopt the following embedding of the string worldsheet into the target spacetime coordinates
\begin{align}
\label{BackStringEmbed}
&x^{\mu}\left(\tau,\sigma\right):t=t\left(\tau\right)\,,\,r=r\left(\tau\right)\,,\,\theta=\pi/2\,,\,\phi=\sigma.
\end{align}In this work, we focus on the dynamics of a circular probe string propagating in the GM/WH spacetime \eqref{GMorWHinEiBI}, with the embedding specified in \eqref{BackStringEmbed}. Under these assumptions, the equation of motion \eqref{EOMsBackEmbedding}, together with the constraints \eqref{RestrictBackEmbedding}, yields the following set of equations
\begin{align}
\label{EOMsBackGMWH}
\ddot{t}\left(\tau\right)\!=\!\ddot{r}\left(\tau\right)\!+\frac{\alpha_{0}^{2}\left(c_{1}r_{0}^{2}\!+\!r(\tau\right)^{2})^{2}\!+\!c_{1}r_{0}^{2}\dot{r}\left(\tau\right)^{2}}{c_{1}r_{0}^{2}r\left(\tau\right)+r\left(\tau\right)^{3}}\!=\!0,\\
\label{RestrictBackGMWH}
\frac{\dot{r}\left(\tau\right)^{2}}{\alpha_{0}^{2}\left(1+\frac{c_{1}r_{0}^{2}}{r(\tau)^{2}}\right)}+r\left(\tau\right)^{2}-\dot{t}\left(\tau\right)^{2}=0.
\end{align}From these equations, the corresponding solutions can be obtained as
\begin{align}
\label{BackStringConfiguraGMWH}
&\bar{t}\left(\tau\right)=\kappa^2 E\tau~,~\bar{r}\left(\tau\right)\!=\!\sqrt{\frac{1}{2}\left(\vert c_{1}r_{0}^{2}\!-\!E^{2}\kappa^{4}\vert\!+\!\left(c_{1}r_{0}^{2}\!+\!E^{2}\kappa^{4}\right)\cos\left(2\alpha_{0}\tau\right)\right)},
\end{align}where the string worldsheet proper time $\tau$ is dimensionless. In addition, to illustrate intuitively how different physical parameters—particularly the sign of $c_1$ and the magnitude of $\alpha_0$—affect the probe string trajectory \eqref{BackStringConfiguraGMWH}, we plot $\bar{r}(\tau)$ for several representative choices of parameters, as shown in Fig.\ref{BackStringGMWH}. It is worth emphasizing that, for the case $c_1=+1$, corresponding to the global monopole spacetime, the parameter $r_0$ must be fixed to zero. Otherwise, $\bar{r}(\tau)$ would contain imaginary values in certain time-domain regions, leading to unphysical behavior. Moreover, as anticipated from the solution \eqref{BackStringConfiguraGMWH}, the string trajectory exhibits periodic motion. From either the explicit solution \eqref{BackStringConfiguraGMWH} or directly from the constraint equation \eqref{RestrictBackGMWH}, the velocity can be evaluated as
\begin{align}
\label{ExpressionDotrBar}
&\dot{\bar{r}}=\alpha_{0}\varepsilon\left(\tau\right)\sqrt{\left(1+\frac{c_{1}r_{0}^{2}}{\bar{r}^{2}}\right)\left(\kappa^{4}E^{2}-\bar{r}^{2}\right)},
\end{align}where the function $\varepsilon(\tau)$ admits the explicit representation in terms of the following Heaviside step functions,
\begin{align}
\nonumber
\varepsilon\left(\tau\right)=&-\sum_{k=0}^{\infty}\Theta\left(\tau-2k\Delta\tau\right)\Theta\left(\left(2k+1\right)\Delta\tau-\tau\right)\\
&+\sum_{k=0}^{\infty}\Theta\left(\tau\!-\!(2k\!+\!1\right)\Delta\tau)\Theta\left(\left(2k+2\right)\Delta\tau\!-\!\tau\right),
\end{align} 
where $\Delta\tau=\frac{\pi}{2\alpha_{0}}$,  denoting the period of the motion. Observing from Fig.\ref{BackStringGMWH},  or directly from the analytic expression \eqref{ExpressionDotrBar}, one can clearly identify a qualitative difference in the behavior of the velocity at $r_0$ between the global monopole and wormhole backgrounds. When the probe string contracts from the asymptotic region at $r_l$ toward $r_0$, its velocity $\bar{r}'\vert_{r_0}$ remains continuous as it passes through $r_0$ in the wormhole case, reflecting the smoothness of the throat geometry. In contrast, for the global monopole background, $\bar{r}'\vert_{r_0}$ exhibits a distinct jump discontinuity, transitioning from $-\alpha_0 \kappa^2 E$ to $\alpha_0 \kappa^2 E$. This discontinuous behavior indicates a qualitatively different effective potential structure near $r_0$, rooted in the nontrivial topological and geometric properties of the global monopole spacetime.

From a physical standpoint, the relevant dynamical degrees of freedom are the transverse fluctuations of the string. In other words, the perturbation $\delta x^{\mu}$ should be projected onto a basis of unit vectors normal to the world-sheet generated by the background configuration $\bar{x}^\mu(\tau,\sigma)$. A systematic construction of such normal vectors and the associated geometric framework can be found in \cite{Larsen:1993mx,Larsen:1998sh,Garriga:1991ts,Guven:1993ex}. For a circular string embedded in a spherically symmetric spacetime, the normal bundle is two-dimensional. The corresponding orthonormal basis can therefore be chosen along two independent polarization directions, which we identify with the radial ($r$) and polar ($\theta$) directions. The polarization index is thus labeled by $\mathtt{i}=r,\theta$, and the transverse perturbation can be decomposed as
\begin{align}
&\delta x^{\mu}\left(\tau,\sigma\right)=\bar{n}_{\left(r\right)}^{\mu}\left(\tau\right)\Phi^{(r)}\left(\tau,\sigma\right)+\bar{n}_{(\theta)}^{\mu}\left(\tau\right)\Phi^{(\theta)}\left(\tau,\sigma\right).
\end{align}Moreover, once the embedding of the circular string given in \eqref{BackStringEmbed} is specified, and the background dynamics are determined through the equations of motion \eqref{EOMsBackGMWH}–\eqref{RestrictBackGMWH} (or equivalently through the explicit solutions \eqref{BackStringConfiguraGMWH}–\eqref{ExpressionDotrBar}), the tangent vectors $\partial\bar{x}^{\nu}/\partial\xi^{\mathtt{A}}$ and the associated orthonormal normal vectors $\bar{n}_{(\mathtt{i})}^{\mu}(\tau)$ can be constructed explicitly. Their concrete expressions take the form

\begin{align}
&\frac{\partial\bar{x}^{\nu}}{\partial\tau}\!\overset{\text{EOMs}}{=\!=\!=}\!\left(\kappa^{2}E,\alpha_{0}\varepsilon(\tau)\sqrt{\left(1\!+\!\frac{c_{1}r_{0}^{2}}{\bar{r}^{2}}\right)\left(\kappa^{4}E^{2}\!-\!\bar{r}^{2}\right)},0,0\right),\\
&\bar{n}_{(r)}^{\mu}\!\!\overset{\text{EOMs}}{=\!=\!=}\!\!\left(\frac{\varepsilon(\tau)}{\bar{r}}\sqrt{\kappa^{4}E^{2}\!-\!\bar{r}^{2}},\frac{\alpha_{0}\kappa^{2}E}{\bar{r}}\!\sqrt{1\!+\!\frac{c_{1}r_{0}^{2}}{\bar{r}^{2}}},0,0\right),\\
&\frac{\partial\bar{x}^{\nu}}{\partial\sigma}=\left(0,0,0,1\right), \\
&\bar{n}_{(\theta)}^{\mu}=\left(0,0,\frac{1}{\bar{r}},0\right).
\end{align}
To derive the quadratic perturbative action for the circular string configuration, one must first compute the relevant geometric objects characterizing the embedding of the world sheet into the target space

\begin{align}
&\Omega_{(\mathtt{i})\,\mathtt{A}\mathtt{B}}=g_{\mu\nu}\bar{n}_{(\mathtt{i})}^{\mu}\frac{\partial\bar{x}^{\alpha}}{\partial\xi^{\mathtt{A}}}\left(\partial_{\alpha}\frac{\partial\bar{x}^{\nu}}{\partial\xi^{\mathtt{B}}}+\Gamma_{\alpha\lambda}^{\nu}\frac{\partial\bar{x}^{\lambda}}{\partial\xi^{\mathtt{B}}}\right)\bigg\vert_{\partial_{r}\to\partial_{\bar{r}(\tau)}}^{r\to\bar{r}(\tau),\theta\to\frac{\pi}{2}},\\
&\mu_{(\mathtt{i})(\mathtt{j})\,\mathtt{A}}=g_{\mu\nu}\bar{n}_{(\mathtt{i})}^{\mu}\frac{\partial\bar{x}^{\alpha}}{\partial\xi^{\mathtt{A}}}\left(\partial_{\alpha}\bar{n}_{(\mathtt{j})}^{\nu}+\Gamma_{\alpha\lambda}^{\nu}\bar{n}_{(\mathtt{j})}^{\lambda}\right)\bigg\vert_{\partial_{r}\to\partial_{\bar{r}(\tau)}}^{r\to\bar{r}(\tau),\theta\to\frac{\pi}{2}},
\end{align}
where $\Omega_{(\mathtt{i}),\mathtt{A}\mathtt{B}}$ denotes the extrinsic curvature of the world sheet, and $\mu_{(\mathtt{i})(\mathtt{j}),\mathtt{A}}$ represents the corresponding surface torsion (normal connection). These quantities encode, respectively, the bending of the world sheet in the ambient spacetime and the twisting of its normal bundle. With these geometric ingredients at hand, the quadratic action governing transverse fluctuations can be expressed in the form
\begin{align}
\nonumber
S_{(2)}&=\frac{1}{2\pi\alpha^{\prime}}\int d\tau d\sigma\sqrt{-h}\,\Phi^{(\mathtt{i})}\left\{\sum_{(\mathtt{k})}h^{\mathtt{A}\mathtt{B}}\left(\delta_{(\mathtt{i})(\mathtt{k})}\nabla_{\mathtt{A}}+\mu_{(\mathtt{i})(\mathtt{k})\mathtt{A}}\right)\right.\\
&\times\left(\delta_{(\mathtt{k})(\mathtt{j})}\nabla_{\mathtt{B}}+\mu_{(\mathtt{k})(\mathtt{j})\mathtt{B}}\right)-\mathcal{V}_{(\mathtt{i})(\mathtt{j})}\Bigg\}\Phi^{(\mathtt{j})},
\end{align}In this expression, $\nabla_{\mathtt{A}}$ is the covariant derivative compatible with the induced world-sheet metric $h_{\mathtt{A}\mathtt{B}}$, ensuring reparametrization covariance at the level of fluctuations. The matrix-valued potential $\mathcal{V}_{(\mathtt{i})(\mathtt{j})}$ is specified by
\begin{align}
\label{PolarPotentialQudraticPer}
\mathcal{V}_{(\mathtt{i})(\mathtt{j})}&=h^{\mathtt{A}\mathtt{B}}R_{\mu\alpha\beta\nu}\partial_{\mathtt{A}}\bar{x}^{\mu}\partial_{\mathtt{B}}\bar{x}^{\nu}\bar{n}_{(\mathtt{i})}^{\alpha}\bar{n}_{(\mathtt{j})}^{\beta}\bigg\vert_{\partial_{r}\to\partial_{\bar{r}(\tau)}}^{r\to\bar{r}(\tau),\theta\to\frac{\pi}{2}}-\frac{2}{G_{~~\mathtt{A}_{1}}^{\mathtt{A}_{1}}}\Omega_{(\mathtt{i})\mathtt{A}\mathtt{B}}\Omega_{(\mathtt{j})}^{~\mathtt{A}\mathtt{B}},
\end{align}which incorporates the effective mass terms generated by both the intrinsic curvature of the world sheet and its extrinsic embedding data. In the expression \eqref{PolarPotentialQudraticPer}, $G_{~~\mathtt{A}_{1}}^{\mathtt{A}_{1}}$ stands for the trace of the induced metric introduced in \eqref{InduceMetric}. The tensor $R_{\mu\alpha\beta\nu}$ denotes the Riemann curvature associated with the background geometry specified by \eqref{BackStringEmbed}. To maintain compatibility with the conventions adopted in the previous subsection, we continue to implement the simplifying replacements $h^{\mathtt{A}\mathtt{B}}\to\eta^{\mathtt{A}\mathtt{B}}$ and $\nabla_{\mathtt{A}}\to\partial_{\mathtt{A}}$. This amounts to adopting the gauge where the world-sheet metric is taken to be flat and the covariant derivative reduces to an ordinary derivative, thereby streamlining the fluctuation analysis without altering the physical content at quadratic order. Specializing to a flat world-sheet metric $\eta_{\mathtt{A}\mathtt{B}}$ and to the GM/WH background given in \eqref{GMorWHinEiBI}, one finds that the extrinsic curvature tensor $\Omega_{(\mathtt{i}),\mathtt{A}\mathtt{B}}$ possesses the following nonvanishing components:
\begin{align}
\label{StringExtrinCurvatureTensor}
&\Omega_{(r),\tau\tau}=\Omega_{(r),\sigma\sigma}=-\alpha_{0}\kappa^{2}E\sqrt{1+\frac{c_{1}r_{0}^{2}}{\bar{r}^{2}}}.
\end{align}Meanwhile, it is straightforward to verify that all components of the surface torsion $\mu_{(\mathtt{i})(\mathtt{j}),\mathtt{A}}$ vanish. Furthermore, upon substituting the above geometric quantities into the potential \eqref{PolarPotentialQudraticPer}, the non-vanishing $\mathcal{V}_{(\mathtt{i})(\mathtt{j})}$ can be found as
\begin{align}
\label{NonVanishPotenRR}
&-\mathcal{V}_{(r)(r)}=\frac{\alpha_{0}^{2}\kappa^{4}E^{2}\left(3c_{1}r_{0}^{2}+2\bar{r}^{2}\right)}{\bar{r}^{4}},\\
\label{NonVanishPotenThetaTheta}
&-\mathcal{V}_{(\theta)(\theta)}=1-\alpha_{0}^{2}-\frac{c_{1}\alpha_{0}^{2}r_{0}^{2}\kappa^{4}E^{2}}{\bar{r}^{4}}.
\end{align}In obtaining Eqs.~\eqref{StringExtrinCurvatureTensor}–\eqref{NonVanishPotenThetaTheta}, we have imposed the constraint $\varepsilon(\tau)^2 = 1$, which considerably simplifies the intermediate expressions while leaving the physical content unchanged. Collecting all contributions, the quadratic effective action governing the string fluctuations can be cast into the form
\begin{align}
\nonumber
S_{(2)}&=\!\frac{1}{2\pi\alpha^{\prime}}\!\int \!\!d\tau d\sigma\left\{\dot{\Phi}_{(r)}^{2}\!-\!\Phi_{(r)}^{\prime2}\!+\frac{\alpha_{0}^{2}\kappa^{4}E^{2}\left(3c_{1}r_{0}^{2}\!+\!2\bar{r}^{2}\right)}{\bar{r}^{4}}\Phi_{(r)}^{2}\right.\\
\label{QuadraticPerString}
&+\dot{\Phi}_{(\theta)}^{2}-\Phi_{(\theta)}^{\prime2}+\left(1-\alpha_{0}^{2}-\frac{c_{1}\alpha_{0}^{2}r_{0}^{2}\kappa^{4}E^{2}}{\bar{r}^{4}}\right)\Phi_{(\theta)}^{2}\Bigg\}.
\end{align}Here and in what follows, an overdot and a prime indicate differentiation with respect to the world-sheet coordinates $\tau$ and $\sigma$, respectively.

\section{Canonical quantization in squeezing formalism \label{SqueeCanoQuantize}}

In this section, we perform the canonical quantization by adopting the squeezing formalism. For a detailed elaboration of the relevant technical details, we refer the reader to Ref.~\cite{Grain:2019vnq}. Because $\sigma=\phi$ is defined on the compact interval $[0,2\pi]$, the fluctuation fields admit a discrete Fourier-mode decomposition of the form
\begin{align}
\label{PerturRadialFourierExpan}
&\Phi_{(r)}\left(\tau,\sigma\right)\!=\!\frac{1}{2\pi}\sum_{n=+2}^{+\infty}\left(\mathrm{R}_{n}(\tau)\text{e}^{\text{i}n\sigma}+\mathrm{R}_{-n}\left(\tau\right)\text{e}^{-\text{i}n\sigma}\right),\\
\label{PerturAngularFourierExpan}
&\Phi_{(\theta)}\left(\tau,\sigma\right)\!=\!\frac{1}{2\pi}\sum_{n=+2}^{+\infty}\left(\Theta_{n}(\tau)\text{e}^{\text{i}n\sigma}+\Theta_{-n}\left(\tau\right)\text{e}^{-\text{i}n\sigma}\right).
\end{align}The reality of the scalar fields $\Phi_{(r)}$ and $\Phi_{(\theta)}$ imposes the constraints $\mathrm{R}_{n}^{\star}(\Theta_{n}^{\star})=\mathrm{R}_{-n}(\Theta_{-n})$, ensuring that the Fourier coefficients satisfy the appropriate complex-conjugation relations. Furthermore, as emphasized in \cite{Garriga:1991ts,Garriga:1991tb,Garriga:1992nm}, the modes with $n=0$ and $n=1$ should be discarded. These modes correspond only to global translations and rigid rotations of the circular string and therefore do not modify its intrinsic shape. Consequently, they do not represent physically meaningful oscillations of the string. Substituting these Fourier expansions into the quadratic-order action \eqref{QuadraticPerString}, we obtain
\begin{align}
\nonumber
S_{(2)}&\!=\!\frac{1}{2\pi^{2}\alpha^{\prime}}\!\!\int \!\! d\tau \!\!\sum_{n=+2}^{+\infty}\left\{\vert\dot{\mathrm{R}}_{n}\vert^{2}\!+\!\left(\frac{\alpha_{0}^{2}\kappa^{4}E^{2}\left(3c_{1}r_{0}^{2}\!+\!2\bar{r}^{2}\right)}{\bar{r}^{4}}\!-\!n^{2}\right)\vert\mathrm{R}_{n}\vert^{2} \right.\\
\label{QuadraActionFourier}
&+\vert\dot{\Theta}_{n}\vert^{2}\!+\left(1\!-\!\alpha_{0}^{2}\!-\frac{c_{1}\alpha_{0}^{2}r_{0}^{2}\kappa^{4}E^{2}}{\bar{r}^{4}}-\!n^{2}\right)\vert\Theta_{n}\vert^{2}\Bigg\},
\end{align}
From the resulting expression \eqref{QuadraActionFourier}, the corresponding conjugate momenta can be defined as
\begin{align}
&\Pi_{l}^{(\mathrm{R})}\left(\tau\right)=\frac{\delta S_{(2)}}{\delta\dot{\mathrm{R}}_{l}^{\star}\left(\tau\right)}=\frac{1}{\pi\alpha^{\prime}}\dot{\mathrm{R}}_{l}\left(\tau\right),\\
&\Pi_{l}^{(\Theta)}\left(\tau\right)=\frac{\delta S_{(2)}}{\delta\dot{\Theta}_{l}^{\star}\left(\tau\right)}=\frac{1}{\pi\alpha^{\prime}}\dot{\Theta}_{l}(\tau).
\end{align}Accordingly, the Hamiltonian of the system is derived as
\begin{align}
\nonumber
H_{(2)}\left(\tau\right)&=\frac{\alpha^{\prime}}{2}\sum_{n=+2}^{\infty}\left(\vert\Pi_{n}^{(\mathrm{R})}\vert^{2}+\vert\Pi_{n}^{(\Theta)}\vert^{2}\right)+\frac{1}{2\pi^{2}\alpha^{\prime}}\sum_{n=+2}^{\infty}\left\{\left(n^{2}\!-\!\frac{\alpha_{0}^{2}\kappa^{4}E^{2}\left(3c_{1}r_{0}^{2}\!+\!2\bar{r}^{2}\right)}{\bar{r}^{4}}\right)\vert\mathrm{R}_{n}\vert^{2} \right.\\ 
\label{QuadraPerHamiltonian}
&+\left(n^{2}-1+\alpha_{0}^{2}+\frac{c_{1}\alpha_{0}^{2}r_{0}^{2}\kappa^{4}E^{2}}{\bar{r}^{4}}\right)\vert\Theta_{n}\vert^{2}\Bigg\}.
\end{align}

After promoting the fluctuations $\Phi_{(r)}, \Pi_{(r)}, \Phi_{(\theta)}, \Pi_{(\theta)}$ to quantum field operators, the corresponding mode expansions can be respectively constructed as
\begin{align}
&\widehat{\Phi}_{(r)}\left(\tau,\sigma\right)=\frac{1}{2\pi}\sum_{n=+2}^{+\infty}\left( {\widehat{\mathrm{R}}_{n}(\tau)}\text{e}^{\text{i}n\sigma}+   {\widehat{\mathrm{R}}^{\dagger}_{n}\left(\tau\right)}\text{e}^{-\text{i}n\sigma} \right),\\
&\hat{\Pi}_{(r)}\left(\tau,\sigma\right)=\frac{1}{2\pi}\sum_{n=+2}^{+\infty}\left(\widehat{\Pi}_{n}^{(\mathrm{R})}\left(\tau\right)\text{e}^{\text{i}n\sigma}+\widehat{\Pi}_{n}^{(\mathrm{R}){\dagger}}\left(\tau\right)\text{e}^{-\text{i}n\sigma} \right),\\
&\widehat{\Phi}_{(\theta)}\left(\tau,\sigma\right)=\frac{1}{2\pi}\sum_{n=+2}^{+\infty}\left(\hat{\Theta}_{n}(\tau)\text{e}^{\text{i}n\sigma}+ \hat{\Theta}_{n}^{\dagger}\left(\tau\right)\text{e}^{-\text{i}n\sigma}\right),\\
&\hat{\Pi}_{(\theta)}\left(\tau,\sigma\right)=\frac{1}{2\pi}\sum_{n=+2}^{+\infty}\left(\widehat{\Pi}_{n}^{(\Theta)}\left(\tau\right)\text{e}^{\text{i}n\sigma}+\widehat{\Pi}_{n}^{(\Theta)\dagger}\left(\tau\right)\text{e}^{-\text{i}n\sigma}\right),
\end{align}    
where 
\begin{align}
     \widehat{\mathrm{R}}_{n}\left(\tau\right)&= \mathcal{R}_{n}\left(\tau\right)\hat{a}_{n}^{(r)}\left(\tau_{0}\right)+\mathcal{R}_{n}^{\star}\left(\tau\right)\hat{a}_{-n}^{(r)\dagger}\left(\tau_{0}\right),\\
     \widehat{\Pi}_{n}^{(\mathrm{R})}\left(\tau\right) &= \frac{\dot{\mathcal{R}}_{n}\left(\tau\right)}{\pi\alpha^{\prime}}\hat{a}_{n}^{(r)}\left(\tau_{0}\right)+\frac{\dot{\mathcal{R}}_{n}^{\star}\left(\tau\right)}{\pi\alpha^{\prime}}\hat{a}_{-n}^{(r)\dagger}\left(\tau_{0}\right),   \\
     \hat{\Theta}_{n}\left(\tau\right)&= \vartheta_{n}\left(\tau\right)\hat{a}_{n}^{(\theta)}\left(\tau_{0}\right)+\vartheta_{n}^{\star}\left(\tau\right)\hat{a}_{-n}^{(\theta)\dagger}\left(\tau_{0}\right),   \\
     \widehat{\Pi}_{n}^{(\Theta)}\left(\tau\right)&= \frac{\dot{\vartheta}_{n}\left(\tau\right)}{\pi\alpha^{\prime}}\hat{a}_{n}^{(\theta)}\left(\tau_{0}\right)+\frac{\dot{\vartheta}_{n}^{\star}\left(\tau\right)}{\pi\alpha^{\prime}}\hat{a}_{-n}^{(\theta)\dagger}\left(\tau_{0}\right).
\end{align}Note that the relation $\hat{\mathcal{O}}_n^\dagger=\hat{\mathcal{O}}_{-n}$ holds for $\hat{\mathcal{O}}=\widehat{\mathrm{R}},\widehat{\Pi}^{(\mathrm{R})},\hat{\Theta},\widehat{\Pi}^{(\Theta)}$. The operators $\hat{a}_{n}^{(\mathtt{i})}(\tau_0),\ \hat{a}_{n}^{(\mathtt{i})\dagger}(\tau_0)$ and $\hat{a}_{-n}^{(\mathtt{i})}(\tau_0),\hat{a}_{-n}^{(\mathtt{i})\dagger}(\tau_0)$ form two independent sets of annihilation and creation operators. Consequently, the only non-vanishing commutation relations are given by
\begin{align}
&2\pi\delta_{nm}\delta^{\mathtt{(i)}\mathtt{(j)}}\!=\left[\hat{a}_{n}^{(\mathtt{i})}\left(\tau_{0}\right),\hat{a}_{m}^{(\mathtt{j})\dagger}\left(\tau_{0}\right)\right]=\left[\hat{a}_{-n}^{(\mathtt{i})}\left(\tau_{0}\right),\hat{a}_{-m}^{(\mathtt{j})\dagger}\left(\tau_{0}\right)\right],
\end{align}
where the indices $\mathtt{i}, \mathtt{j}$ denote the polarization labels associated with the $r$ and $\theta$ directions. Here, $\tau_0$ indicates that these operators act on the vacuum state defined at the initial time. In implementing the standard canonical quantization procedure, one must impose the equal-time commutation relations
\begin{align}
\nonumber
\text{i}\delta\left(\sigma-\sigma^{\prime}\right)&=\left[\hat{\Phi}_{(r)}\left(\tau,\sigma\right),\hat{\Pi}_{(r)}\left(\tau,\sigma^{\prime}\right)\right]\\
&=\left[\hat{\Phi}_{(\theta)}\left(\tau,\sigma\right),\hat{\Pi}_{(\theta)}\left(\tau,\sigma^{\prime}\right)\right].
\end{align}From these relations, the corresponding Wronskian normalization condition for the mode functions can be derived as
\begin{align}
\nonumber
\text{i}\pi\alpha^{\prime}&=\mathcal{R}_{n}\left(\tau\right)\dot{\mathcal{R}}_{n}^{\star}\left(\tau\right)-\mathcal{R}_{n}^{\star}\left(\tau\right)\dot{\mathcal{R}}_{n}\left(\tau\right)\\
\label{WronskianMode}
&=\vartheta_{n}\left(\tau\right)\dot{\vartheta}_{n}^{\star}\left(\tau\right)-\vartheta_{n}^{\star}\left(\tau\right)\dot{\vartheta}_{n}\left(\tau\right)\, , \, n\geq 2.
\end{align}Furthermore, the time evolution of the mode functions is governed by the following equations of motion:
\begin{align}
\label{EOMsRadial}
\frac{d^{2}\mathcal{R}_{n}\left(\tau\right)}{d\tau^{2}}-\left(\frac{\alpha_{0}^{2}\kappa^{4}E^{2}\left(3c_{1}r_{0}^{2}\!+\!2\bar{r}^{2}\right)}{\bar{r}^{4}}-\!n^{2}\right)\mathcal{R}_{n}\left(\tau\right)&=0,\\
\label{EOMsAngular}
\frac{d^{2}\vartheta_{n}\left(\tau\right)}{d\tau^{2}}-\left(1\!-\!\alpha_{0}^{2}\!-\frac{c_{1}\alpha_{0}^{2}r_{0}^{2}\kappa^{4}E^{2}}{\bar{r}^{4}}-\!n^{2}\right)\vartheta_{n}(\tau)&=0.
\end{align}

By transforming the phase-space field variables into the helicity basis \cite{Grain:2019vnq}, we implement a linear canonical transformation that diagonalizes the system at the level of independent polarization modes. In this basis, a time-dependent annihilation operator can be defined as
\begin{align}
\label{LinearCanonical}
&\hat{a}_{\pm n}^{(r)}\left(\tau\right)=\frac{1}{\sqrt{2\alpha^{\prime}}}\hat{\mathrm{R}}_{\pm n}\left(\tau\right)+\text{i}\sqrt{\frac{\alpha^{\prime}}{2}}\hat{\Pi}_{\pm n}^{(\mathrm{R})}\left(\tau\right),
\end{align}while the corresponding creation operator is obtained via Hermitian conjugation of $\hat{a}_{\pm n}^{(r)}(\tau)$. An entirely analogous construction applies to the operators $\hat{a}_{\pm n}^{(\theta)}(\tau)$ and $\hat{a}_{\mp n}^{(\theta)\dagger}(\tau)$ associated with the angular polarization modes. Within this framework, the initial conditions for the mode functions $\mathcal{R}_{n}(\tau)$ and $\vartheta_{n}(\tau)$ are fixed by the operator definitions above, and can be expressed as
\begin{align}
\label{InitialConditionModes}
&\vartheta_{n}\left(\tau_{0}\right)\!=\!\frac{\text{i}}{\pi}\dot{\vartheta}_{n}\left(\tau_{0}\right)\!=\!\mathcal{R}_{n}\left(\tau_{0}\right)\!=\!\frac{\text{i}}{\pi}\dot{\mathcal{R}}_{n}\left(\tau_{0}\right)\!=\!\sqrt{\frac{\alpha^{\prime}}{2}}.
\end{align}It is important to note that this initial condition is consistent with the Wronskian normalization \eqref{WronskianMode}. The combination of this normalization condition and the chosen initial condition is not arbitrary; rather, it ensures a well-defined vacuum state at the initial time $\tau_0$. As will be shown later, this choice indeed guarantees that the mean particle number vanishes at $\tau_0$. Furthermore, upon substituting these time-dependent operators back into \eqref{QuadraPerHamiltonian}, the quadratic Hamiltonian operator, $\hat{\mathcal{H}}^{\text{(2)}}=\hat{\mathcal{H}}_{(r)}^{\text{(2)}}+\hat{\mathcal{H}}_{(\theta)}^{\text{(2)}}$, can be expressed in terms of the time-dependent creation and annihilation operators as
\begin{align}
\nonumber
\hat{\mathcal{H}}_{(r)}^{\text{(2)}}&=\frac{1}{4\pi^{2}}\sum_{n=2}^{\infty}\left\{2\left(n^{2}\!-\!\frac{\alpha_{0}^{2'/\kappa^{4}E^{2}\left(3c_{1}r_{0}^{2}\!+\!2\bar{r}^{2}\right)}}{\bar{r}^{4}}+\pi^{2}\right)\hat{\mathcal{K}}_{z,n}^{(r)}\right.\\
\label{HamilQuadraticPerInr}
&+\left(n^{2}\!-\!\frac{\alpha_{0}^{2}\kappa^{4}E^{2}(3c_{1}r_{0}^{2}\!+\!2\bar{r}^{2})}{\bar{r}^{4}}-\pi^{2}\right)\left(\text{i}\hat{\mathcal{K}}_{-,n}^{(r)}-\text{i}\hat{\mathcal{K}}_{+,n}^{(r)}\right)\Bigg\},\\
\nonumber
\hat{\mathcal{H}}_{(\theta)}^{\text{(2)}}&=\frac{1}{4\pi^{2}}\sum_{n=2}^{\infty}\left\{2\left(n^{2}\!-\!1\!+\!\alpha_{0}^{2}\!+\frac{c_{1}\alpha_{0}^{2}r_{0}^{2}\kappa^{4}E^{2}}{\bar{r}^{4}}+\!\pi^{2}\right)\hat{\mathcal{K}}_{z,n}^{(\theta)}\right.\\
\label{HamilQuadraticPerInTheta}
&+\left(n^{2}\!-\!1\!+\!\alpha_{0}^{2}\!+\frac{c_{1}\alpha_{0}^{2}r_{0}^{2}\kappa^{4}E^{2}}{\bar{r}^{4}}-\!\pi^{2}\right)\left(\text{i}\hat{\mathcal{K}}_{-,n}^{(\theta)}\!-\!\text{i}\hat{\mathcal{K}}_{+,n}^{(\theta)}\right)\Bigg\},
\end{align}
where $\hat{\mathcal{K}}_{\pm,n}^{(\mathtt{i})}$ and $\hat{\mathcal{K}}_{z,n}^{(\mathtt{i})}$ are the generators of the $su(1,1)$ algebra. These generators admit a realization in terms of bilinear combinations of creation and annihilation operators, known as the Schwinger representation, given by
\begin{align}
\label{su11ShiftOpe}
&\hat{\mathcal{K}}_{+,n}^{(\mathtt{i})}\!=\!\text{i}\hat{a}_{n}^{(\mathtt{i})\dagger}\left(\tau\right)\hat{a}_{-n}^{(\mathtt{i})\dagger}\left(\tau\right)\, , \, \hat{\mathcal{K}}_{-,n}^{(\mathtt{i})}\!=\!-\text{i}\hat{a}_{-n}^{(\mathtt{i})}(\tau)\hat{a}_{n}^{(\mathtt{i})}\left(\tau\right),\\
\label{su11WeightOpe}
&\hat{\mathcal{K}}_{z,n}^{(\mathtt{i})}\!=\!\frac{1}{2}\left(\hat{a}_{n}^{(\mathtt{i})\dagger}\hat{a}_{n}^{(\mathtt{i})}+\hat{a}_{-n}^{(\mathtt{i})}\hat{a}_{-n}^{(\mathtt{i})\dagger}\right).
\end{align}
From \eqref{su11ShiftOpe}–\eqref{su11WeightOpe}, one can readily verify that these operators satisfy the standard $su(1,1)$ commutation relations,
\begin{align}
&[\hat{\mathcal{K}}_{+}^{(\mathtt{i})},\hat{\mathcal{K}}_{-}^{(\mathtt{i})}]\!=\!-2\hat{\mathcal{K}}_{z}^{(\mathtt{i})}\,,\,[\hat{\mathcal{K}}_{z}^{(\mathtt{i})},\hat{\mathcal{K}}_{\pm}^{(\mathtt{i})}]\!=\!\pm\hat{\mathcal{K}}_{\pm}^{(\mathtt{i})},
\end{align}where we have introduced the notation $\hat{\mathcal{K}}_{\pm;z}^{(\mathtt{i})}=\frac{1}{2\pi}\sum_{n=1}\hat{\mathcal{K}}_{\pm;z,n}^{(\mathtt{i})}$ for brief. In principle, given the quadratic Hamiltonian \eqref{HamilQuadraticPerInr}–\eqref{HamilQuadraticPerInTheta}, the time-evolution operator $\hat{\mathcal{U}}(\tau,\tau_{0})$ can be formally constructed as
\begin{align}
\label{DefineTimeEvoluOperator}
&\hat{\mathcal{U}}^{(\mathtt{i})}\left(\tau,\tau_{0}\right)=\mathcal{T}\exp\left(-\text{i}\int_{\tau_{0}}^{\tau}d\tilde{\tau}\,\hat{\mathcal{H}}_{(\mathtt{i})}^{\text{(2)}}(\tilde{\tau})\right).
\end{align}However, in practical computations, directly expanding the exponential of operators composed of creation and annihilation operators is technically cumbersome, as it requires the use of the Zassenhaus formula together with the evaluation of a large number of nested commutators. Instead, it is far more efficient to exploit the underlying symmetry structure: the $su(1,1)$ Lie algebra provides a systematic and tractable framework for constructing the time-evolution operator. Exploiting the left-polar decomposition of a generic $SU(1,1)$ group element \cite{Barnett,Puri:2001,Grain:2019vnq,Li:2021kfq}, one can express it in the form
\begin{align}
\label{TimeEvolutionOpeDecompose}
&\hat{\mathcal{U}}^{(\mathtt{i})}\left(\tau,\tau_{0}\right)\!=\! \hat{\mathcal{S}}^{(\mathtt{i})} \hat{\mathcal{R}}^{(\mathtt{i})},
\end{align}
where 
\begin{align}
   \hat{\mathcal{S}}^{(\mathtt{i})}&=\exp\left(\frac{1}{2\pi}\sum^\infty_{n=2}\left(\xi_{n}^{(\mathtt{i})}\left(\tau\right)\hat{\mathcal{K}}_{+,n}^{(\mathtt{i})}\left(\tau_{0}\right)-\xi_{n}^{(\mathtt{i})\star}\left(\tau\right)\hat{\mathcal{K}}_{-,n}^{(\mathtt{i})}\left(\tau_{0}\right)\right)\right),  \\
    \hat{\mathcal{R}}^{(\mathtt{i})}&=\exp\left(\frac{\text{i}}{\pi}\sum^\infty_{n=2}\varpi_{n}^{(\mathtt{i})}\left(\tau\right)\hat{\mathcal{K}}_{z,n}^{(\mathtt{i})}\left(\tau_{0}\right)\right).
\end{align}Here $\xi_{n}^{(\mathtt{i})}$ is conveniently parametrized as $\xi_{m}^{(\mathtt{i})}(\tau)=-\text{i}\gamma_{m}^{(\mathtt{i})}(\tau)\text{e}^{2\text{i}\varphi_{m}^{(\mathtt{i})}(\tau)}$, which is the time-dependent parameter. Within the squeezed-state formalism, $\gamma_{m}^{(\mathtt{i})}(\tau)$ and $\varphi_{m}^{(\mathtt{i})}(\tau)$ correspond to the squeezing amplitude and squeezing phase, respectively, while $\varpi_{m}^{(\mathtt{i})}(\tau)$ denotes the rotation angle. Accordingly, the operators $\hat{\mathcal{S}}\big(\gamma(\tau),\varphi(\tau)\big)$ and $\hat{\mathcal{R}}\big(\varpi(\tau)\big)$ represent the two-mode squeezing operator and the rotation operator. Furthermore, by invoking standard operator-ordering theorems \cite{Barnett,Puri:2001}, the squeezing operator can be factorized into a disentangled form as

\begin{align}
\nonumber
\hat{\mathcal{S}}^{(\mathtt{i})}\left(\gamma,\varphi\right)\!&=\!\exp\left\{\frac{1}{2\pi}\sum^\infty_{m=2}\left(\text{e}^{2\text{i}\varphi_{m}^{(\mathtt{i})}}\tanh\left(\gamma_{m}^{(\mathtt{i})}\right)\hat{a}_{m}^{(\mathtt{i})\dagger}\left(\tau_{0}\right)\hat{a}_{-m}^{(\mathtt{i})\dagger}\left(\tau_{0}\right)\right)\right\}\\
\nonumber
&\times\exp\left\{\frac{1}{2\pi}\sum^\infty_{m=2}\left(-\ln(\cosh(\gamma_{m}^{(\mathtt{i})}))\left(\hat{a}_{-m}^{(\mathtt{i})}\hat{a}_{-m}^{(\mathtt{i})\dagger}+\hat{a}_{m}^{(\mathtt{i})\dagger}\hat{a}_{m}^{(\mathtt{i})}\right)\right)\right\}\\
&\times\exp\left\{\frac{1}{2\pi}\sum^\infty_{m=2}\left(\!-\!\text{e}^{-2\text{i}\varphi_{m}^{(\mathtt{i})}}\tanh\left(\gamma_{m}^{(\mathtt{i})}\right)\hat{a}_{-m}^{(\mathtt{i})}\left(\tau_{0}\right)\hat{a}_{m}^{(\mathtt{i})}\left(\tau_{0}\right)\right)\right\}.
\end{align}
For a detailed derivation of the operator-ordering theorems associated with the Lie group $SU(1,1)$, we refer the reader to Refs.~\cite{Barnett,Puri:2001,Grain:2019vnq,Martin:2015qta,Li:2021kfq}. On the other hand, in addition to the linear canonical transformation~\eqref{LinearCanonical}, the time evolution of the creation and annihilation operators can be constructed with the aid of the evolution operator $\hat{\mathcal{U}}^{(\mathtt{i})}(\tau,\tau_{0})$. In particular, the operators at an arbitrary time $\tau$, namely $\hat{a}_{n}(\tau)$ and $\hat{a}_{n}^{\dagger}(\tau)$, are given by
\begin{align}
\label{ConstruCreationOpeAnyTau}
&\hat{a}_{\pm n}^{(\mathtt{i})}\left(\tau\right)=\hat{\mathcal{R}}^{(\mathtt{i})\dagger}(\varpi)\hat{\mathcal{S}}^{(\mathtt{i})\dagger}\left(\gamma,\varphi\right)\hat{a}_{\pm n}^{(\mathtt{i})}\left(\tau_{0}\right)\hat{\mathcal{S}}^{(\mathtt{i})}\left(\gamma,\varphi\right)\hat{\mathcal{R}}^{(\mathtt{i})}\left(\varpi\right),
\end{align}
where no summation over the repeated polarization index $\mathtt{i}$ is implied. A completely analogous construction applies to $\hat{a}_{\pm n}^{(\mathtt{i})\dagger}(\tau)$, following the same formalism as in~\eqref{ConstruCreationOpeAnyTau}. By repeatedly applying the commutator expansion formula
$\text{e}^{\hat{B}}\hat{A}\text{e}^{-\hat{B}}=\hat{A}+[\hat{B},\hat{A}]+\frac{1}{2!}[\hat{B},[\hat{B},\hat{A}]]+\dots$,
one eventually obtains
\begin{align}
\label{CreationOpeAnyTauSqueeze}
&\hat{a}_{n}^{(\mathtt{i})}\left(\tau\right)=\cosh\left(\gamma_{n}^{(\mathtt{i})}(\tau)\right)\text{e}^{\text{i}\varpi_{n}^{(\mathtt{i})}(\tau)}\hat{a}_{n}^{(\mathtt{i})}\left(\tau_{0}\right)+\text{e}^{-\text{i}\varpi_{n}^{(\mathtt{i})}\left(\tau\right)}\text{e}^{2\text{i}\varphi_{n}^{(\mathtt{i})}\left(\tau\right)}\sinh\left(\gamma_{n}^{(\mathtt{i})}(\tau)\right)\hat{a}_{-n}^{(\mathtt{i})\dagger}\left(\tau_{0}\right),\\
\label{AnnihiOpeAnyTauSqueeze}
&\hat{a}_{-n}^{(\mathtt{i})\dagger}\left(\tau\right)=\cosh\left(\gamma_{n}^{(\mathtt{i})}\left(\tau\right)\right)\text{e}^{-\text{i}\varpi_{n}^{(\mathtt{i})}(\tau)}\hat{a}_{-n}^{(\mathtt{i})\dagger}(\tau_{0})+\text{e}^{\text{i}\varpi_{n}^{(\mathtt{i})}(\tau)}\text{e}^{-2\text{i}\varphi_{n}^{(\mathtt{i})}(\tau)}\sinh\left(\gamma_{n}^{(\mathtt{i})}(\tau)\right)\hat{a}_{n}^{(\mathtt{i})}\left(\tau_{0}\right).
\end{align}By matching the expressions in \eqref{CreationOpeAnyTauSqueeze}--\eqref{AnnihiOpeAnyTauSqueeze} with those derived from the linear canonical transformation~\eqref{LinearCanonical}, we arrive at

\begin{align}
\label{SqueeAngleToModeFunc}
&\cos\left(2\varphi_{n}^{(r)}\right)\!=\!\frac{\text{Re}\left\{\!\left(\mathcal{R}_{n}\left(\tau\right)\!+\!\frac{\text{i}}{\pi}\!\dot{\mathcal{R}}_{n}(\tau)\big)\big(\mathcal{R}_{n}^{\star}(\tau)\!+\!\frac{\text{i}}{\pi}\!\dot{\mathcal{R}}_{n}^{\star}\left(\tau\right)\right)\!\right\}}{\bigg\vert\mathcal{R}_{n}(\tau)\!+\!\frac{\text{i}}{\pi}\!\dot{\mathcal{R}}_{n}\left(\tau\right)\bigg\vert\cdot\bigg\vert\mathcal{R}_{n}\left(\tau\right)\!-\!\frac{\text{i}}{\pi}\!\dot{\mathcal{R}}_{n}\left(\tau\right)\bigg\vert},\\
\label{SqueeAmpliToModeFunc}
&\sinh^{2}\left(\gamma_{n}^{(r)}\right)=\frac{1}{2\alpha^{\prime}}\bigg\vert\mathcal{R}_{n}(\tau)-\frac{\text{i}}{\pi}\dot{\mathcal{R}}_{n}(\tau)\bigg\vert^{2},\\
\label{RotationFactorToModeFunc}
&\cos\left(\varpi_{n}^{(r)}\right)=\frac{\text{Re}\left(\mathcal{R}_{n}\left(\tau\right)+\frac{\text{i}}{\pi}\dot{\mathcal{R}}_{n}(\tau)\right)}{\bigg\vert\mathcal{R}_{n}(\tau)+\frac{\text{i}}{\pi}\dot{\mathcal{R}}_{n}(\tau)\bigg\vert}.
\end{align}
\begin{figure}[t]
 \begin{center}
    \includegraphics[scale=0.325]{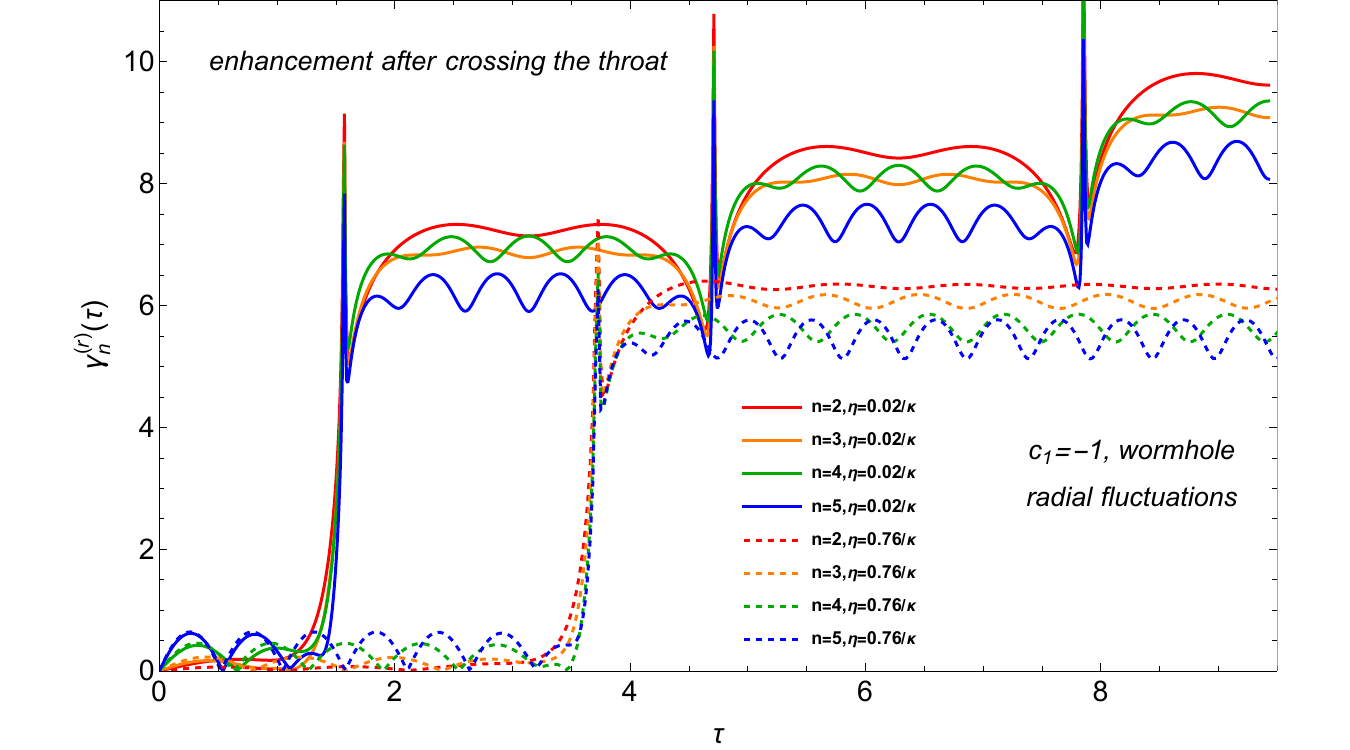}
    \includegraphics[scale=0.325]{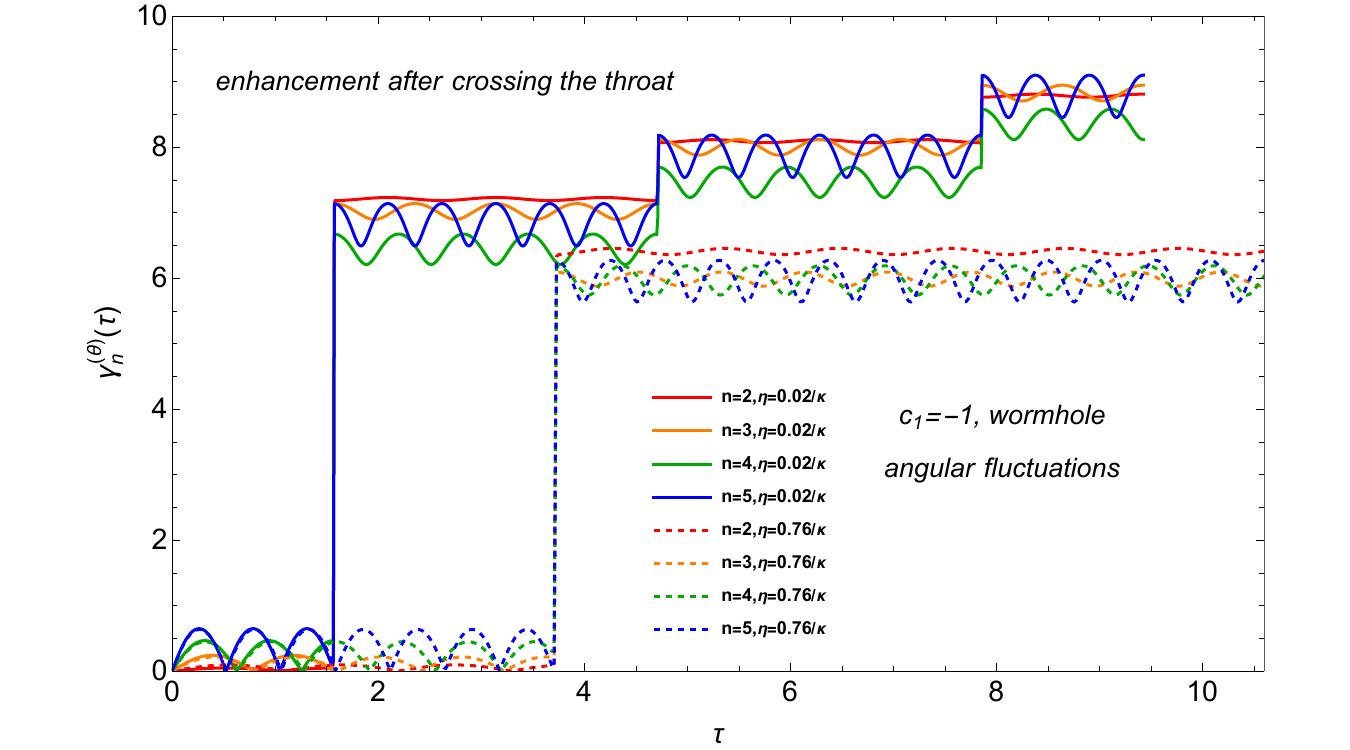}\\
    \includegraphics[scale=0.325]{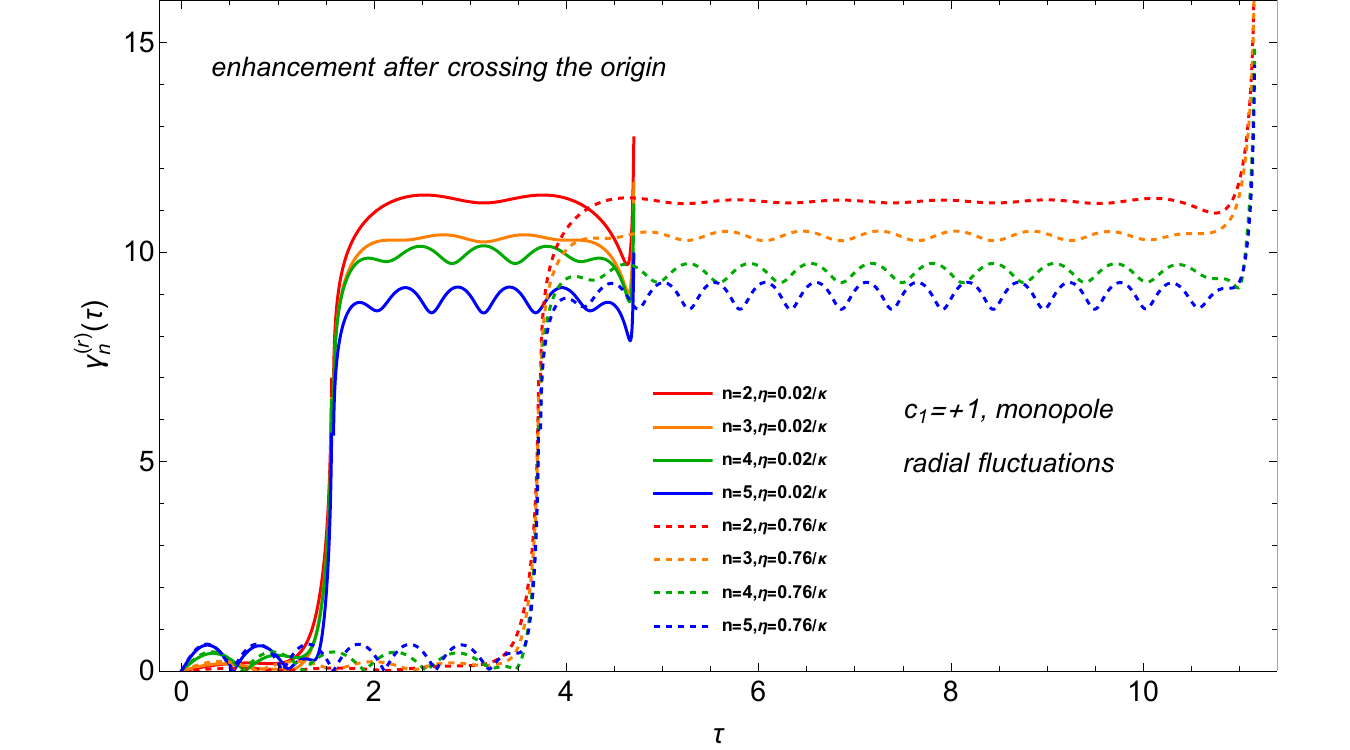}
     \includegraphics[scale=0.325]{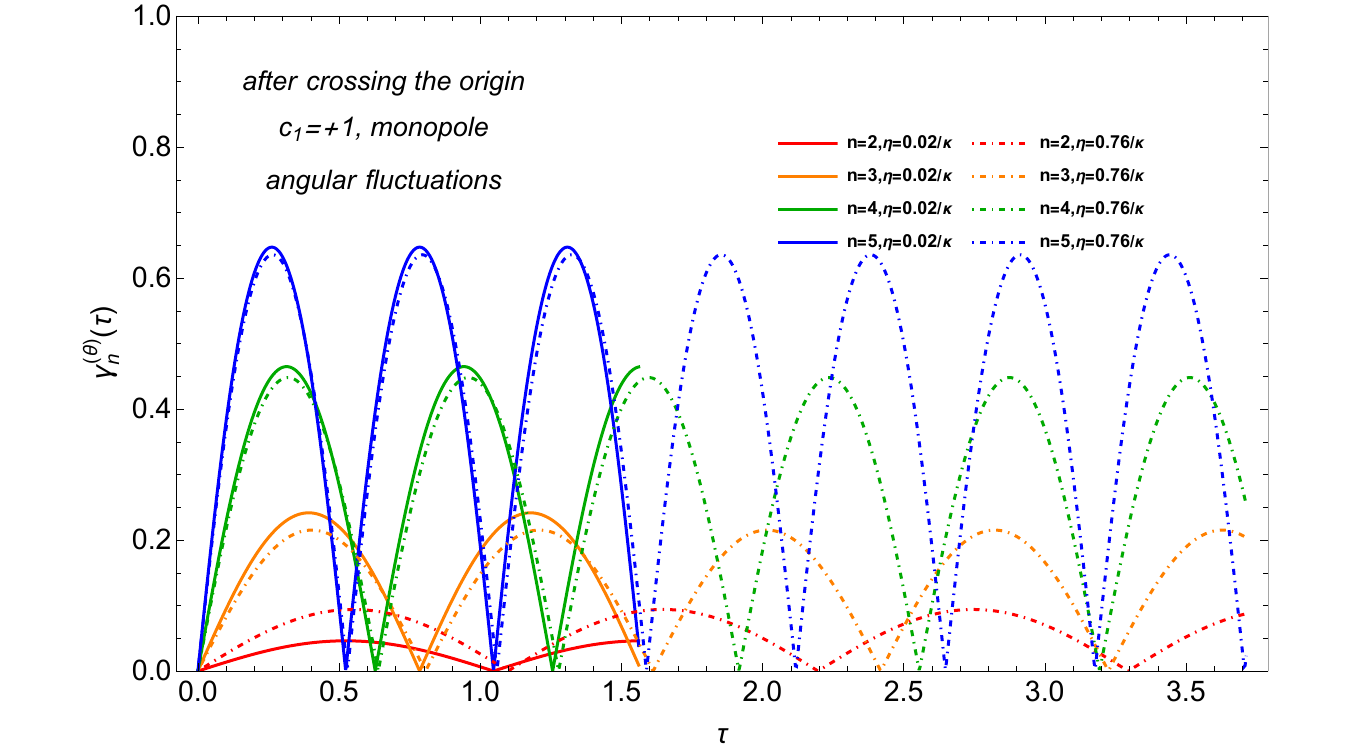}
 \caption{Time evolution of $\gamma^{(\mathtt{i})}_n (\tau)$, which characterizes the amplitudes of the two-mode quantum states generated by the quantum fluctuations of the probe string propagating in the global monopole and monopole wormhole background spacetimes.}
 \label{SqueezAmplitudeForGMWH}
 \end{center}
 \end{figure}
An identical set of expressions is obtained for the case of fluctuations in the angular polarization sector, upon making the replacement $\mathcal{R}_n \to \vartheta_{n}$. Given the initial conditions in \eqref{InitialConditionModes}, it follows straightforwardly that $\gamma_{n}^{(\mathtt{i})}(\tau_{0})=\varpi_{n}^{(\mathtt{i})}(\tau_{0})=\varphi_{n}^{(\mathtt{i})}(\tau_{0})=0$. After numerically solving the equations of motion~\eqref{EOMsRadial}-\eqref{EOMsAngular} and subsequently substituting the corresponding solutions into the squeezing parameters, in particular the amplitude $\gamma_{n}^{(\mathtt{i})}(\tau)$ as given in~\eqref{SqueeAmpliToModeFunc}, we obtain the results shown in Fig.~\ref{SqueezAmplitudeForGMWH}. The accuracy of the numerical solutions is monitored using the normalization conditions in \eqref{WronskianMode}. In generating Fig.~\ref{SqueezAmplitudeForGMWH}, we choose the parameter $L_0$ appropriately such that $r_{0}=L_{0}\kappa\eta$ is held fixed for different values of $\eta$ in the wormhole case $c_1=-1$. Physically, this corresponds to comparing wormhole configurations with different $\eta$ while keeping the throat radius the same. Note that, in the monopole wormhole case with $c_1=-1$, the solution reduces to the Ellis–Bronnikov wormhole~\cite{Ellis:1973yv,Bronnikov:1973fh} in the limit $\eta \to 0$.

As illustrated in Fig.~\ref{SqueezAmplitudeForGMWH}, a comparison with the Ellis–Bronnikov case (represented by the solid curves) shows that the presence of the monopole, characterized by the parameter $\eta$, leads to a significant suppression of the amplitudes of the quantum states excited by quantum fluctuations (as indicated by the dashed curves). This suppression can be attributed to the $\eta$-dependent deformation of the background geometry, which modifies the effective mode evolution and consequently reduces the efficiency of particle production. We also note that, due to the signature of the spacetime metric, the parameter $\eta$ is constrained to satisfy $\eta<1/\kappa$. Within this allowed range, $\eta$ thus effectively acts as a suppressing parameter for the strength of quantum fluctuations in the monopole wormhole configuration. In contrast, for the pure monopole case with $c_1=+1$, the corresponding amplitudes of the quantum states do not exhibit a significant dependence on $\eta$. This qualitative difference between the two backgrounds may hint at a deeper connection, possibly related to the ER=EPR conjecture \cite{Maldacena:2013xja,Susskind:2014yaa}. Motivated by this observation, in the following we focus on the entanglement entropy associated with these quantum fluctuations, with particular emphasis on its dependence on $\eta$ in the global monopole and monopole wormhole backgrounds.
\begin{figure}[t]
 \begin{center}
 \includegraphics[scale=0.325]{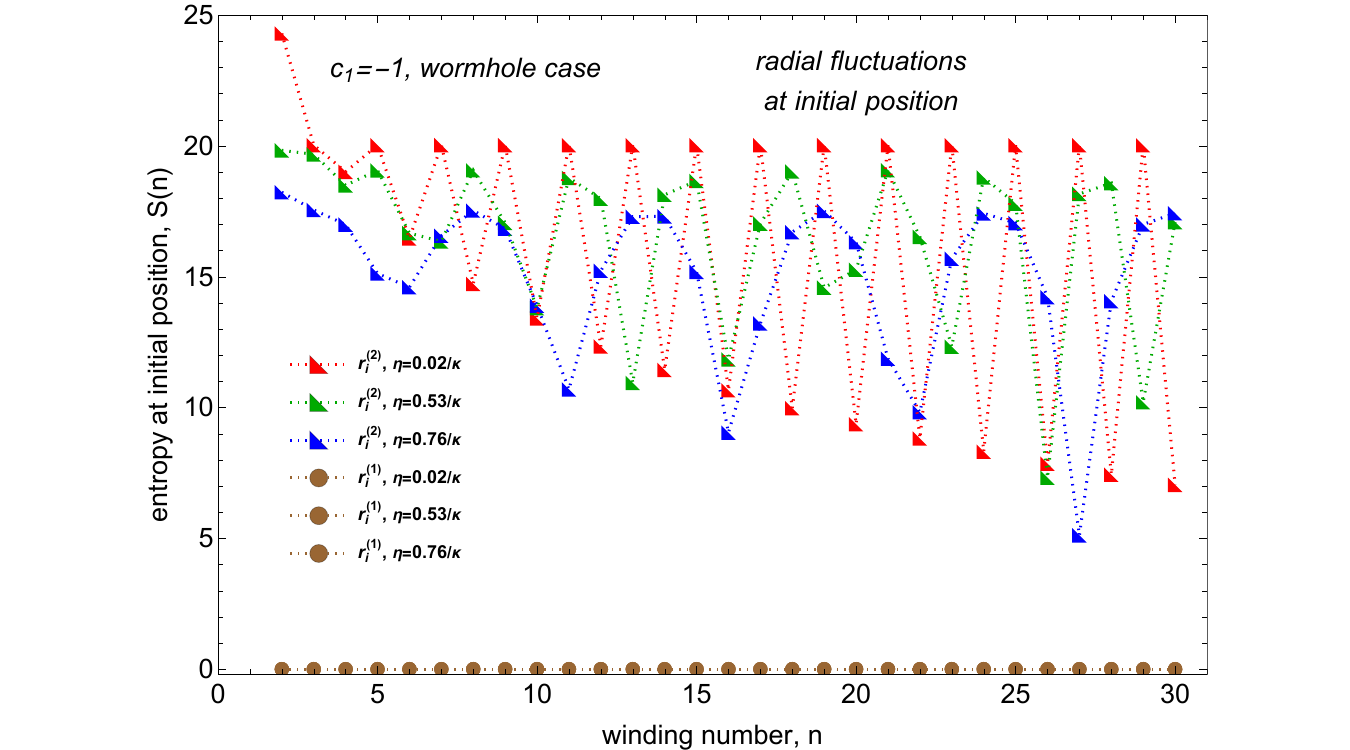}
  \includegraphics[scale=0.325]{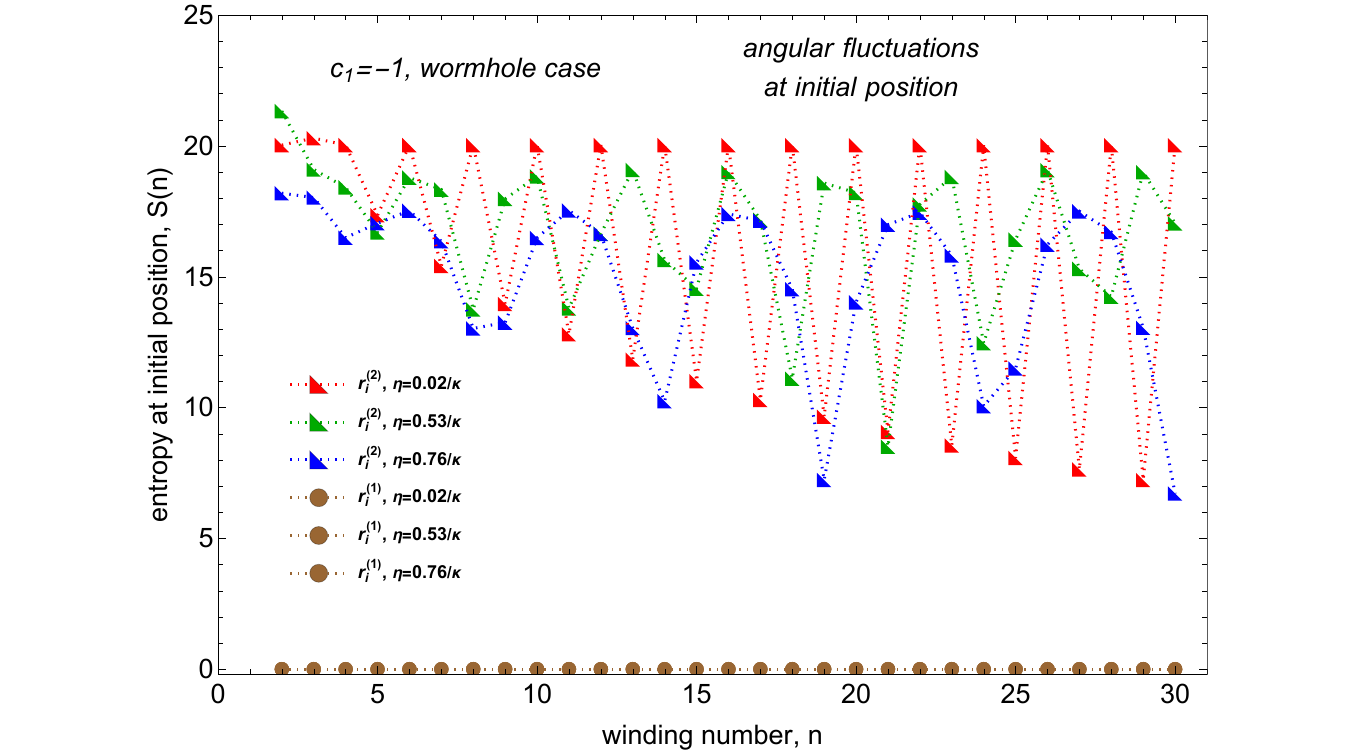}\\
    \includegraphics[scale=0.325]{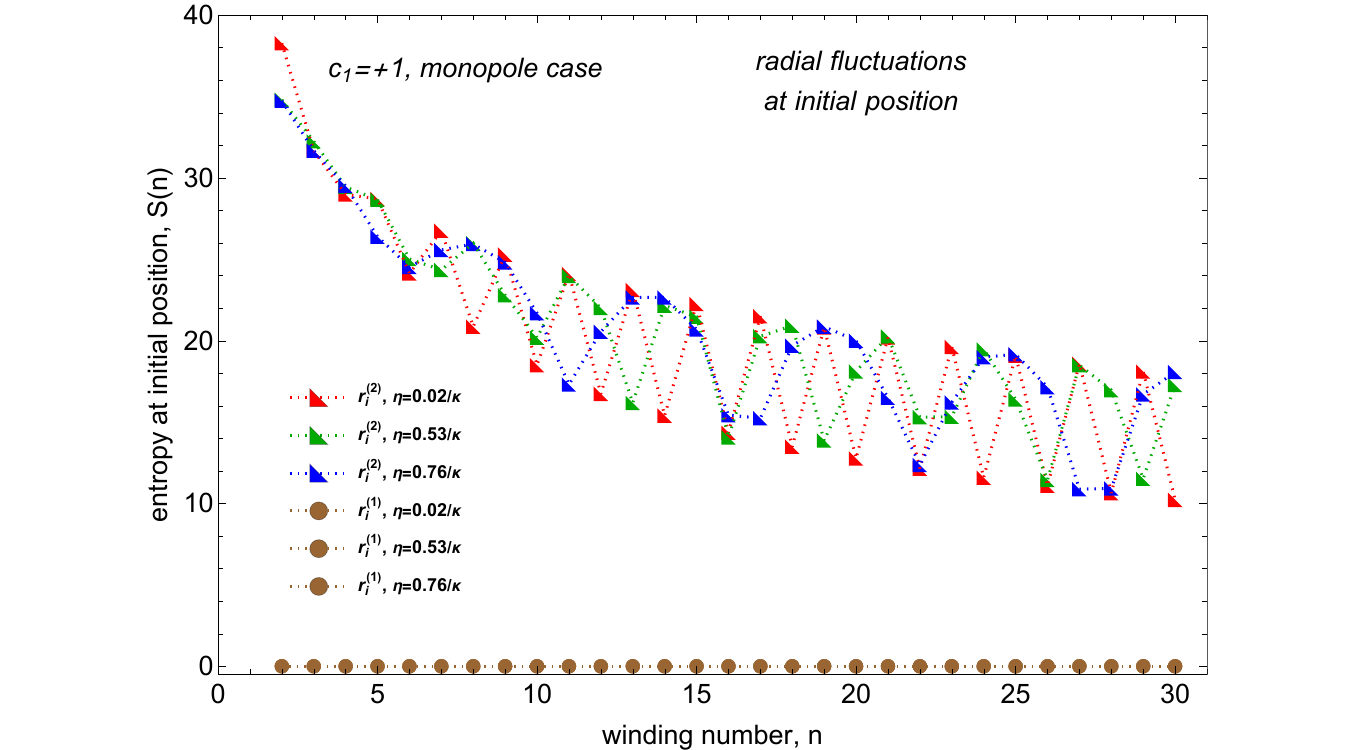}
    \includegraphics[scale=0.325]{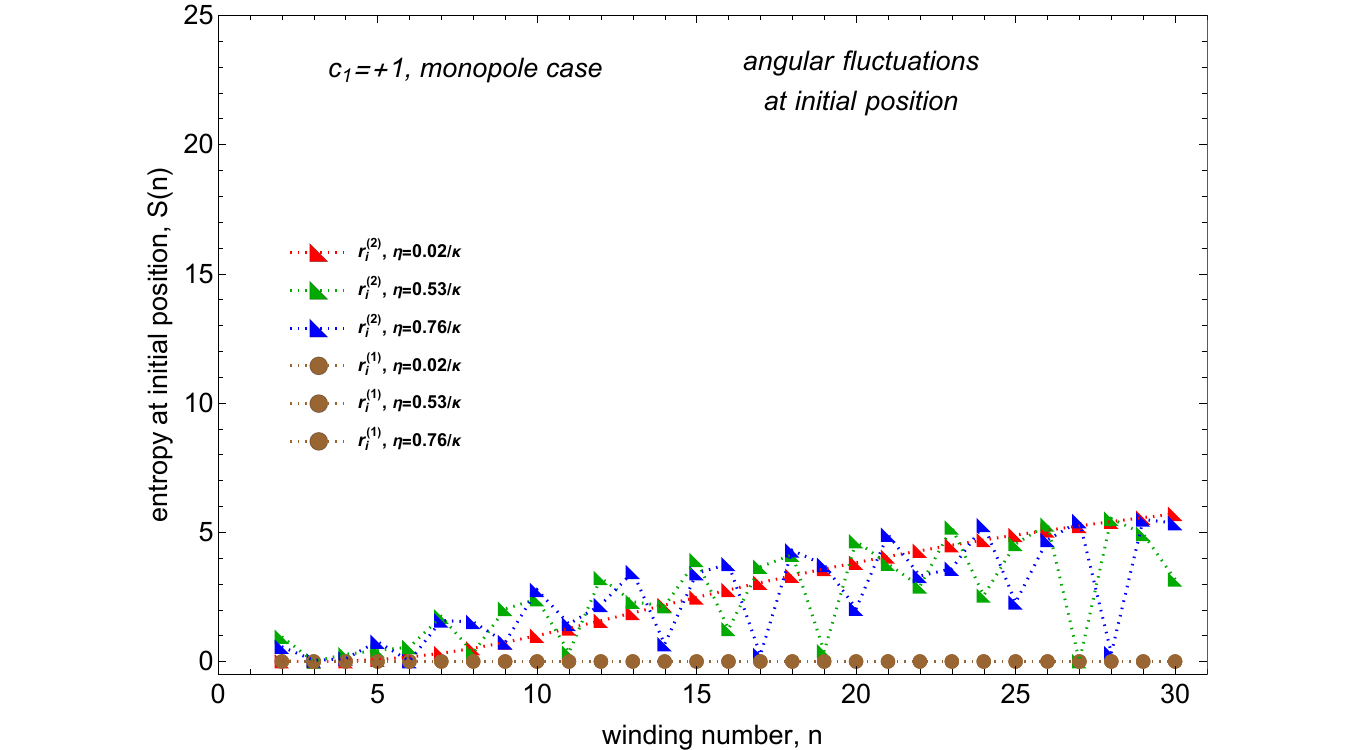}
 \caption{Enhancement of the entanglement entropy spectrum as a function of the winding number after a single passage through the origin in the monopole case, defined as $\Delta S_n = S_n\vert_{r_i^{(2)}} - S_n\vert_{r_i^{(1)}}$. Here, $r_i^{(1)}$ and $r_i^{(2)}$ denote, respectively, the initial position of the probe string at $\tau=\tau_0$, i.e., $r_i^{(1)}=\bar{r}(\tau_0)$, and the position at the same spatial point after one complete cycle of motion, namely $r_i^{(2)}=\bar{r}(\tau_0+\Delta \tau)$, corresponding to the string traveling from the initial position to the monopole origin and returning back.}
 \label{EntroVsWindNumGMWH}
 \end{center}
 \end{figure}

\section{Two-mode quantum state and the relevant Von Neumann entropy \label{QuantumStateAndEntropy}}

\begin{figure}
 \begin{center}
    \includegraphics[scale=0.55]{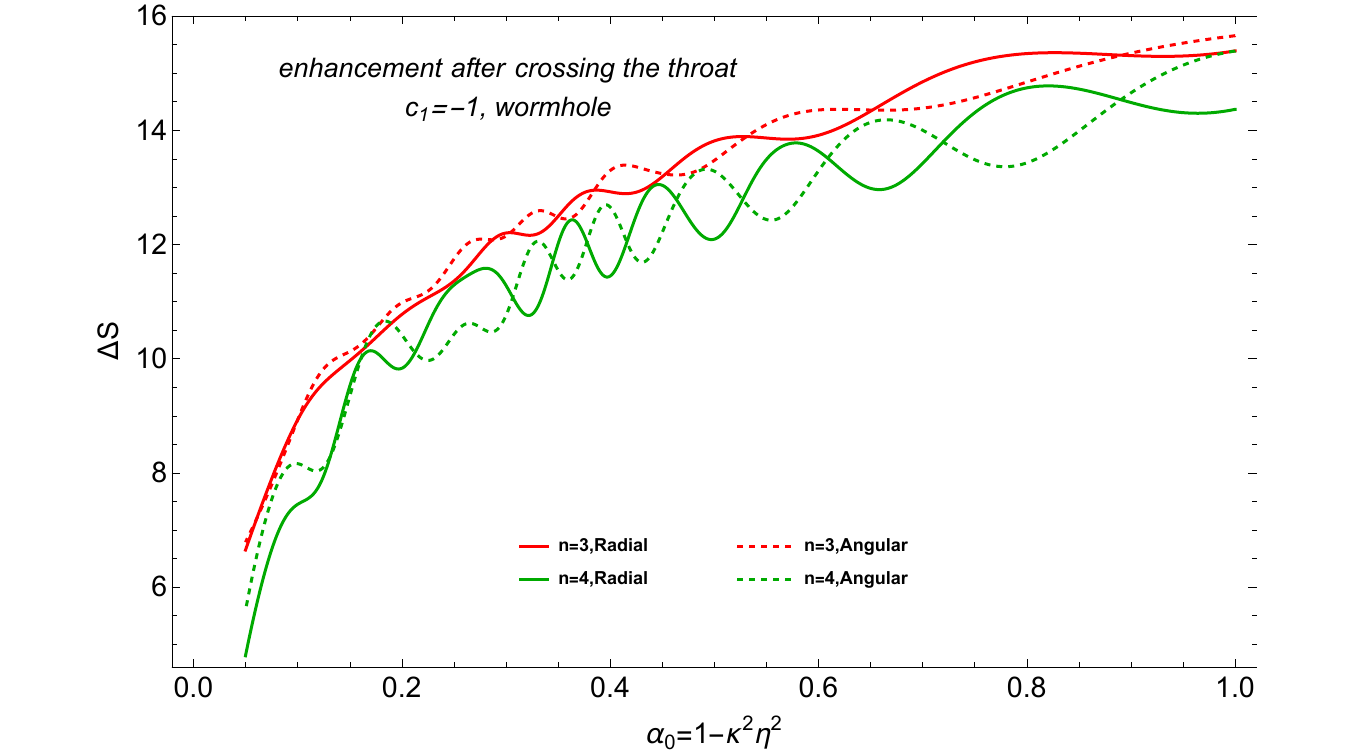}\\
    \includegraphics[scale=0.55]{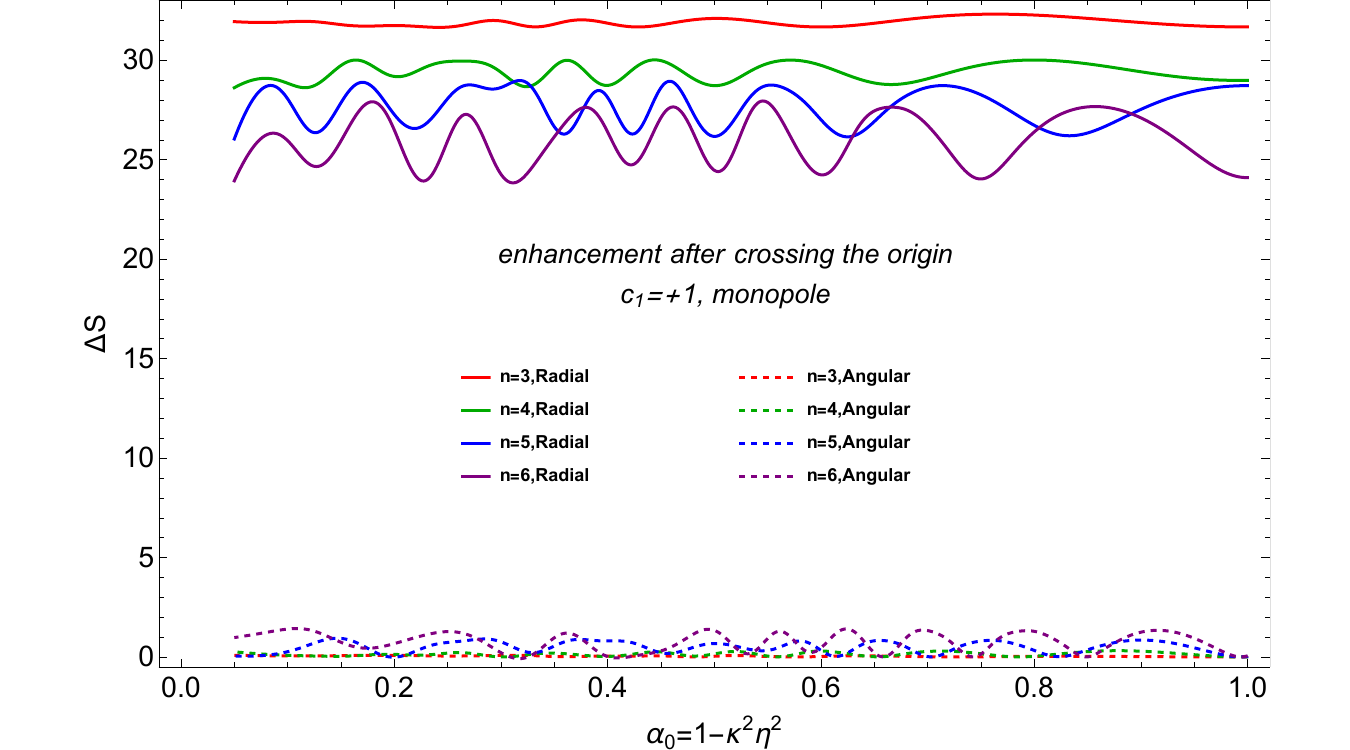}
 \caption{Variation of the quantities $\Delta S_n = S_n\vert_{r_i^{(2)}} - S_n\vert_{r_i^{(1)}}$, corresponding to Fig.~\ref{EntroVsWindNumGMWH}, as a function of the deficit angle factor $\alpha_0=1-8\pi G \eta^2$, for several fixed winding numbers chosen as representative examples. In the upper panel, which corresponds to the monopole wormhole configuration, we display the behavior of $\Delta S_n$ as a function of $\eta$ only for winding numbers $n=2,3$, while the cases with $n=4,5$ are not shown since their corresponding curves nearly overlap with those for $n=2,3$. We further note that for larger winding numbers, the dependence of $\Delta S_n$ on $\eta$ exhibits the same qualitative behavior as in the representative cases presented here, differing only by the presence of additional oscillatory features.}
 \label{EnhanceEntropyCompareWHandGM}
 \end{center}
 \end{figure}

With the aid of the time-evolution operator $\hat{\mathcal{U}}^{(\mathtt{i})}(\tau,\tau_{0})$, the corresponding two-mode quantum state can be constructed as
\begin{align}
\nonumber
\vert\Psi_{\gamma,\varphi,\varpi}^{(\mathtt{i})}\left(\tau\right)\rangle &=\prod_{n=2}^{\infty}\hat{\mathcal{U}}^{(\mathtt{i})}\left(\tau;\tau_{0}\right)\vert\tilde{0}_{n},\tilde{0}_{-n}\rangle_{\tau_{0}}=\prod_{n=2}^{\infty}\hat{\mathcal{S}}^{(\mathtt{i})}\left(\gamma,\varphi\right)\hat{\mathcal{R}}^{(\mathtt{i})}\left(\varpi\right)\vert\tilde{0}_{n},\tilde{0}_{-n}\rangle_{\tau_{0}}\\
\label{TwoModeQuantumStateAnyTau}
&=\prod_{m=2}^{\infty}\frac{\text{e}^{\text{i}\varpi_{m}^{(\mathtt{i})}\left(\tau\right)}}{\cosh\left(\gamma_{m}^{(\mathtt{i})}\left(\tau\right)\right)}\sum_{\tilde{n}=0}^{\infty}\left(\text{e}^{2\text{i}\varphi_{m}^{(\mathtt{i})}\left(\tau\right)}\tanh\left(\gamma_{m}^{(\mathtt{i})}\left(\tau\right)\right)\right)^{\tilde{n}}\big\vert\tilde{n}_{m},\tilde{n}_{-m}\rangle_{\tau_{0}},
\end{align}
where $\tilde{m}$ and $\tilde{n}$ represent the occupation numbers of the quantum state. They are introduced to distinguish these quantum numbers from the winding numbers $m$ and $n$ appearing in the Fourier expansions \eqref{PerturRadialFourierExpan}–\eqref{PerturAngularFourierExpan}. For the two-mode quantum states defined above, the associated density matrix takes the form \cite{Ball:2005xa,Martin-Martinez:2012chf,Pierini:2018wki}
\begin{align}
\hat{\rho}^{(\mathtt{i})}\left(n,-n\right)&=\frac{1}{\cosh\left(\gamma_{n}^{(\mathtt{i})}\left(\tau\right)\right)^{2}}\sum_{\tilde{n},\tilde{m}=0}^{\infty}\left(\tanh(\gamma_{n}^{(\mathtt{i})}(\tau))\right)^{\tilde{m}+\tilde{n}} \text{e}^{2\text{i}\left(\tilde{n}-\tilde{m}\right)\varphi_{n}^{(\mathtt{i})}(\tau)}~\vert\tilde{n}_{n},\tilde{n}_{-n}\rangle\langle\tilde{m}_{n},\tilde{m}_{-n}\vert.
\end{align}
where, for notational simplicity, the reference time label $\tau_{0}$ in the basis states $\vert\tilde{n}_{n},\tilde{n}_{-n}\rangle_{\tau_0}$ has been suppressed. The degree of particle–antiparticle entanglement can be characterized through the reduced density matrix. In particular, starting from the full density matrix $\hat{\rho}^{(\mathtt{i})}(n,-n)$, one obtains the reduced density matrix $\hat{\rho}^{(\mathtt{i})}(n)$ by tracing over the degrees of freedom associated with the negative winding number mode “$-n$”. This procedure yields
\begin{align}
&\hat{\rho}^{(\mathtt{i})}\left(n\right)=\sum_{\tilde{l}=0}^{\infty}\langle\tilde{l}_{-n}\vert\hat{\rho}^{(\mathtt{i})}\left(n,-n\right)\vert\tilde{l}_{-n}\rangle=\frac{1}{\cosh\left(\gamma_{n}^{(\mathtt{i})}\left(\tau\right)\right)^{2}}\sum_{\tilde{l}=0}^{\infty}\left(\tanh\left(\gamma_{n}^{(\mathtt{i})}\left(\tau\right)\right)\right)^{2\tilde{l}}\vert\tilde{l}_{n}\rangle\langle\tilde{l}_{n}\vert.
\end{align}Once the reduced density matrix $\hat{\rho}^{(\mathtt{i})}(n)$ is obtained, the corresponding von Neumann entropy can be computed as
\begin{align}
\nonumber
S\left[\hat{\rho}^{(\mathtt{i})}(n)\right]&\!=\!-\text{Tr}\left\{\hat{\rho}^{(\mathtt{i})}(n)\log_{2}\big(\hat{\rho}^{(\mathtt{i})}\left(n\right)\big)\right\}\! \notag \\
&=\!-\sum_{\tilde{l}=0}^{\infty}\frac{\left(\tanh(\gamma_{n}^{(\mathtt{i})}\left(\tau\right))\right)^{2\tilde{l}}}{\cosh\left(\gamma_{n}^{(\mathtt{i})}\left(\tau\right)\right)^{2}}\log_{2}\left(\frac{\left(\tanh\left(\gamma_{n}^{(\mathtt{i})}\left(\tau\right)\right)\right)^{2\tilde{l}}}{\cosh\left(\gamma_{n}^{(\mathtt{i})}\left(\tau\right)\right)^{2}}\right)\\
\label{VonNeuMannEntropy}
&=\left(1+\mathcal{N}_{n}^{(\mathtt{i})}\left(\tau\right)\right)\log_{2}\left(1+\mathcal{N}_{n}^{(\mathtt{i})}\left(\tau\right)\right)-\mathcal{N}_{n}^{(\mathtt{i})}\left(\tau\right)\log_{2}\mathcal{N}_{n}^{(\mathtt{i})}\left(\tau\right),
\end{align}where, in deriving this expression, we have employed the definition of the mean occupation number
\begin{align}
\nonumber
\mathcal{N}_{n}^{(\mathtt{i})}\left(\tau\right)&={}_{\tau_{0}}\langle\tilde{0}_{n},\tilde{0}_{-n}\vert\frac{1}{2\pi}\sum^\infty_{m=2}\hat{a}_{m}^{(\mathtt{i})\dagger}\left(\tau\right)\hat{a}_{n}^{(\mathtt{i})}\left(\tau\right)\vert\tilde{0}_{n},\tilde{0}_{-n}\rangle_{\tau_{0}}\\
\label{ParticleNum}
&=\sinh\left(\gamma_{n}^{(\mathtt{i})}\left(\tau\right)\right)^{2}.
\end{align}
In fact, the expression for the particle number in \eqref{ParticleNum} corresponds to a thermal distribution whenever $\gamma_{n}^{(\mathtt{i})} \neq 0$. Substituting the amplitudes $\gamma_{n}^{(r,\theta)}$, written in terms of the mode functions as in \eqref{SqueeAmpliToModeFunc}, into the entropy formula \eqref{VonNeuMannEntropy}, one can compute the entanglement spectrum as a function of the winding number. In particular, by evaluating the entropy at several representative time slices and for different values of $\eta$, we obtain the dependence of the entanglement entropy on the winding number. The resulting spectra are illustrated in Fig.~\ref{EntroVsWindNumGMWH}. It can be clearly observed that whenever the probe string passes through the throat in the wormhole case, or through the origin in the monopole case, the entanglement entropy undergoes a pronounced enhancement for each winding number mode. In addition, the magnitude of this enhancement depends sensitively on the value of $\eta$. This dependence becomes especially prominent in the monopole wormhole configuration, where the variation in the entropy growth for different $\eta$ values is particularly significant. To further quantify this behavior, we present Fig.~\ref{EnhanceEntropyCompareWHandGM}, which illustrates the dependence of the entropy increase $\Delta S_{n}$ on $\eta$ after the probe string completes one full cycle of its periodic motion.

In the case of the monopole wormhole, as the probe string moves from a large-$r$ region toward the wormhole throat and subsequently returns to large $r$, thereby completing one cycle of its periodic motion, the entanglement entropy exhibits a clear enhancement with increasing deficit angle factor $\alpha_0 = 1 - 8\pi G \eta^2$, as illustrated in the upper panel of Fig.~\ref{EnhanceEntropyCompareWHandGM}. The parameter $\alpha_0$ characterizes the effective size of the physical spatial section after removing the deficit angle associated with the global monopole, and therefore provides a measure of the remaining geometric scale of the spacetime. It is also worth emphasizing that, after one complete cycle of the periodic motion in the monopole wormhole background, the quantum fluctuations induced in the radial and angular polarization sectors remain of comparable magnitude. This feature is reflected both in the enhancement of the amplitude of the two-mode quantum state, as shown in Fig.~\ref{SqueezAmplitudeForGMWH}, and in the corresponding increase of the entanglement entropy, displayed in Fig.~\ref{EntroVsWindNumGMWH}. Since the latter quantity is a functional of the former, the comparable growth in amplitude naturally translates into a similar enhancement in the entanglement entropy for the two polarization channels. By contrast, the situation differs markedly in the global monopole spacetime. On the one hand, the increase of the entanglement entropy after one cycle of periodic motion shows an extremely weak dependence on the deficit angle factor $\alpha_0$, remaining nearly constant over the range considered. On the other hand, the quantum fluctuations generated in the radial and angular polarization directions differ significantly in magnitude. This pronounced hierarchy appears both in the squeezing amplitude $\gamma_n^{(\mathtt{i})}(\tau)$ and in the associated entanglement entropy $S_n^{(\mathtt{i})}(\tau)$, as illustrated in Fig.~\ref{EntroVsWindNumGMWH} and Fig.~\ref{EnhanceEntropyCompareWHandGM}. These results indicate that, although wormholes and topological defects such as monopoles share several common geometric features, for example both typically involve violations of the classical energy conditions, their impacts on quantum probes can differ significantly. In particular, when quantum observables such as entanglement are considered, the two types of spacetimes lead to markedly different behaviors. For the case in which the global monopole spacetime serves as the background geometry, the variation of the entanglement entropy shows little sensitivity to the defect parameter that characterizes the spacetime structure. By contrast, when the monopole wormhole acts as the background spacetime, the entanglement entropy becomes strongly dependent on the parameter $\eta$, which measures the strength of the underlying topological defect. In other words, the entanglement dynamics in the wormhole geometry responds sensitively to changes in the defect structure of the spacetime. This behavior is, to some extent, reminiscent of the general spirit of the ER=EPR conjecture. We stress, however, that the entanglement considered here is not an intrinsic quantum entanglement of the classical wormhole geometry itself. The background GM/WH spacetime is treated as a fixed classical geometry, while the von Neumann entropy computed above measures the particle--antiparticle entanglement of the quantized fluctuation modes of the circular string probe. The topological charge and the wormhole throat affect this entropy only through their influence on the effective mode evolution and the resulting two-mode squeezed state of the probe fluctuations.

\section{Conclusion and discussion}\label{ConAndDiscu}

In this work, we begin by reviewing a class of global monopole/wormhole (GM/WH) spacetime solutions arising in Eddington-inspired Born–Infeld (EiBI) gravity. We then show that a probe cosmic string embedded in this GM/WH background admits periodic motion along its classical trajectory. Building on the classical trajectory of this probe string, we derive the corresponding quadratic-order perturbative action that describes small fluctuations around the background configuration, including perturbations in both the radial and angular polarization directions. Within this perturbative framework, we then perform the canonical quantization of the system using the squeezing formalism. By applying the time-evolution operator to the Fock space constructed from the initial vacuum state, we obtain the corresponding two-mode quantum states that encode the quantum excitations of the string fluctuations. Since such states are intrinsically entangled, the corresponding Von Neumann entropy provides a natural measure of the particle–antiparticle entanglement generated by the quantum fluctuations of the probe string during its periodic motion, particularly when the string passes through the wormhole throat or the monopole origin. We have also clarified how the GM/WH geometry used in this work is connected with known limiting cases. In particular, the positive branch corresponds to the global monopole sector, while the negative branch describes a topologically charged wormhole which reduces, in the zero-deficit limit with fixed throat radius, to the Ellis--Bronnikov wormhole. Therefore, the entanglement spectrum obtained here may be regarded as a quantum-probe extension of previous analyses based on scalar fields, geodesics, and gravitational lensing.

Our results reveal a clear qualitative distinction between topological defects with and without a wormhole structure. In the monopole wormhole background, the entanglement entropy of the probe string increases significantly with the deficit angle factor $\alpha_0 = 1 - 8\pi G \eta^2$, indicating that the quantum correlations are sensitive to the bulk geometric scale. This behavior is broadly consistent with the spirit of the ER=EPR conjecture. By contrast, in the global monopole spacetime without a wormhole, the entanglement generated by the quantum fluctuations of the probe string shows only a very weak dependence on the bulk size, suggesting that the presence of a wormhole geometry plays a crucial role in linking spacetime structure to quantum entanglement.

More generally, the differences between topological defect spacetimes and strong-gravity objects such as wormholes and black holes have been extensively investigated from classical perspectives, for example through the geodesic motion of massive or massless probe particles and through gravitational lensing. By contrast, studies based on genuinely quantum probes remain relatively limited. In certain asymptotically AdS geometries, the holographic dictionary within the framework of the AdS/CFT correspondence allows quantum observables to be inferred from geometric quantities in the dual bulk spacetime. For non-AdS backgrounds, however, such explorations are much less developed. The cosmic string probe combined with the squeezed-state quantization approach adopted in this work provides a relatively general framework for introducing quantum probes, such as entanglement, into non-AdS spacetimes and for extracting physical insights from their dynamical evolution.

In the present study we focus on the particle–antiparticle entanglement associated with the quantum fluctuations of a circular cosmic string in two asymptotically flat but geometrically distinct backgrounds, namely the global monopole spacetime and the monopole wormhole spacetime. As a natural extension, this framework could be applied to more complicated spacetime geometries, such as black-bounce Schwarzschild spacetimes with different bounce parameters \cite{Simpson:2018tsi} or various objects in AdS backgrounds or some novel dynamical spacetimes \cite{Zeng:2018pzk,Zeng:2023ueq,McVittie:1933zz,Bochicchio:2010df,Faraoni:2005it,Maeda:2009tk,Faraoni:2008tx,Roman:1992xj}. In such contexts, it would be interesting to employ quantum probes to distinguish among geometries including pure AdS spacetime, AdS solitons, and AdS black holes or wormholes. One could further compare the entanglement measures obtained directly from quantum-information methods with geometric quantities suggested by the AdS/CFT correspondence, thereby providing an additional perspective on the relation between spacetime geometry and quantum correlations. 

Another intriguing direction arises from an analogy with studies of holographic thermalization \cite{Albash:2010mv,Balasubramanian:2011ur}. In that context, holographic entanglement entropy, computed through extremal bulk surfaces anchored on the boundary subsystem, serves as an important diagnostic tool for tracking the approach of a nonequilibrium state toward thermal equilibrium during gravitational collapse in the bulk. Motivated by this idea, one may speculate that the entanglement generated by the quantum fluctuations of a circular cosmic string could play a similar diagnostic role. In particular, by embedding such a quantum probe into a dynamical spacetime undergoing gravitational collapse and black hole formation, one could use the resulting entanglement evolution as a probe of the thermalization process associated with black hole formation, especially in spacetimes that are not asymptotically AdS where the standard holographic dictionary is absent. This perspective may provide a complementary way to investigate non-equilibrium gravitational dynamics from the viewpoint of quantum probes.

\subsection*{Some comments on decoherence effects}

It is important to first distinguish the present fixed-background probe analysis from dynamical discussions of wormhole rupture or deconstruction in ER=EPR-inspired settings. In holographic contexts, loss of entanglement or decoherence has been argued to be associated with the pinching-off, breaking, or deconstruction of an Einstein--Rosen bridge~\cite{VanRaamsdonk2010,Verlinde2021,Murdia2023}. By contrast, our calculation does not include gravitational backreaction, topology change, or a dynamical evolution of the background geometry. Therefore, any possible decoherence mentioned here should be understood as decoherence among the probe fluctuation modes or polarization sectors, rather than as a literal rupture of the topologically charged wormhole.

Nevertheless, the present results contain an observation that is qualitatively suggestive in this direction. In the monopole wormhole background, increasing the topological defect parameter $\eta$ tends to suppress the particle--antiparticle entanglement of the circular-string probe. This suppression can be seen from the upper panels of Fig.~\ref{SqueezAmplitudeForGMWH} and Fig.~\ref{EntroVsWindNumGMWH}, and more directly from Fig.~\ref{EnhanceEntropyCompareWHandGM}. Since a larger $\eta$ corresponds to a stronger solid-angle deficit, or equivalently to a smaller deficit-angle factor $\alpha_0=1-8\pi G\eta^2$, this behavior may be interpreted heuristically as reflecting a reduction of the effective spatial scale associated with the wormhole geometry. In this limited diagnostic sense, the suppression of probe-mode entanglement is reminiscent of the broader idea that decoherence or loss of entanglement can be related to the degradation of an Einstein--Rosen bridge. We emphasize, however, that this analogy does not imply an actual rupture of the fixed background wormhole in the present probe approximation.

Finally, a noteworthy feature emerges from comparing the entanglement spectra shown in Fig.~\ref{EntroVsWindNumGMWH}. When only radial fluctuations are considered, the entanglement entropy in the monopole background exceeds that in the monopole wormhole background. In contrast, when angular fluctuations are included, the trend is reversed. This observation suggests that the relative behavior of the two backgrounds, particularly in the monopole case, is highly sensitive to the polarization sector under consideration. A plausible explanation is that the present analysis is restricted to quadratic-order perturbations, within which radial and angular modes remain decoupled and evolve independently. Once higher-order contributions, such as cubic interactions, are incorporated, couplings between the two sectors are expected to emerge. These interactions may induce mode mixing and effectively act as channels for decoherence, allowing information exchange between different polarization modes. As a consequence, the entanglement entropies associated with the two sectors may tend to equilibrate, reducing the disparity observed at the quadratic level. A systematic investigation of higher-order perturbative effects and the associated decoherence dynamics is left for future work.

Future investigations based on open quantum system techniques or influence-functional approaches may provide a more complete understanding of decoherence effects induced by nontrivial spacetime topology. In particular, it would be worthwhile to investigate whether the topological sector of the curved background may effectively behave as an environment for the quantum probe, thereby inducing geometry-dependent suppression of quantum correlations, in qualitative analogy with recent discussions on wormhole-entanglement interplay and holographic decoherence.

\section{Acknowledge}

Ai-chen Li was supported by funding from the China Scholarship Council (CSC) with  Grant No. 202008620074.
Xin-Fei Li is supported by NSFC with Grant No. 12565008, Youth Program of Natural Science Foundation of Guangxi with Grant No. 2021GXNSFBA075049 and Doctor Start-up Foundation of Guangxi University of Science and Technology with Grant No. 19Z21.

\bibliographystyle{unsrt}
\bibliography{bibliography}

%\cite{Nascimento:2019qor}
%\bibitem{Nascimento:2019qor}
%J.~R.~Nascimento, G.~J.~Olmo, P.~J.~Porf{\'\i}rio, A.~Y.~Petrov and A.~R.~Soares,
%``Nonlinear $\sigma$-models in the Eddington-inspired Born-Infeld Gravity,''
%Phys. Rev. D \textbf{101}, no.6, 064043 (2020)
%doi:10.1103/PhysRevD.101.064043
%[arXiv:1912.10779 [hep-th]].
%37 citations counted in INSPIRE as of 04 Mar 2026

\end{document}